\documentclass[11pt]{article}
\pdfoutput=1 

\usepackage{jcappub} 

\usepackage[T1]{fontenc} 
\usepackage{natbib}
\usepackage{amsmath,amssymb,bm}
\usepackage{enumitem,units,multirow}
\usepackage{hyperref}

\newcommand{\ie}{{\it i.e.\ }}

\newcommand{\eg}{{\it e.g.\ }}

\newcommand{\py}{\small\ttfamily}

\title{\boldmath {\sc FAST-PT} II: an algorithm to calculate convolution integrals of general tensor quantities in cosmological perturbation theory}

\author{Xiao Fang,}
\author{Jonathan A. Blazek,}
\author{Joseph E. McEwen, and}
\author{Christopher M. Hirata}

\affiliation{Center for Cosmology and AstroParticle Physics, Department of Physics, The Ohio State University, 191 W Woodruff Ave, Columbus OH 43210, USA}

\emailAdd{fang.307@osu.edu}
\emailAdd{blazek@berkeley.edu}
\emailAdd{mcewen.24@osu.edu}
\emailAdd{hirata.10@osu.edu}

\abstract{
Cosmological perturbation theory is a powerful tool to predict the statistics of large-scale structure in the weakly non-linear regime, but even at 1-loop order it results in computationally expensive mode-coupling integrals. Here we present a fast algorithm for computing 1-loop power spectra of quantities that depend on the observer's orientation, thereby generalizing the {\py FAST-PT} framework (McEwen {\it et al.}, 2016) that was originally developed for scalars such as the matter density. This algorithm works for an arbitrary input power spectrum and substantially reduces the time required for numerical evaluation. We apply the algorithm to four examples: intrinsic alignments of galaxies in the tidal torque model; the Ostriker-Vishniac effect; the secondary CMB polarization due to baryon flows; and the 1-loop matter power spectrum in redshift space. Code implementing this algorithm and these applications is publicly available at \url{https://github.com/JoeMcEwen/FAST-PT}.}

\newcommand{\Pl}{P_\text{lin}} 

\newcommand{\dq}[1]{\frac{d^3 \bm{q}_{#1}}{(2\pi)^3}}

\begin{document}
\maketitle
\flushbottom

\clearpage

\section{Introduction}
\label{sec:intro} 

Observational cosmology has entered a new era of precision measurement. Current and upcoming surveys \cite{Levi:2013gra,BOSS2013AJ....145...10D,EUCLID2011arXiv1110.3193L,WFIRST2013arXiv1305.5422S,Abbott:2016ktf} are enabling us to probe large-scale structure in more detail and over larger volumes, and hence to better constrain the underlying cosmological model. A parallel effort is underway to understand the astrophysical effects that are both signals and contaminants in these measurements. For example, weak gravitational lensing has become a powerful and direct probe of the dark matter distribution \cite{Bartelmann:1999yn,Mellier:1998pk}, but it also suffers from systematic uncertainties, such as galaxy intrinsic alignments (IA), which must be mitigated in order to make use of high-precision measurements. Similarly, connecting observable tracers (\eg in spectroscopic surveys) with the underlying dark matter requires a description of the bias relationship \cite{Seljak:2004sj,McDonald:2006mx,2009JCAP...08..020M,Baldauf:2011bh,2012JCAP...03..004S} and the effect of redshift-space distortions (RSDs) \cite{1987MNRAS.227....1K,Scoccimarro:2004tg,Taruya:2010mx}. Developments in CMB measurements provide another illustration, as the range of observables has expanded from early initial detections of temperature anisotropies by COBE \cite{1992SvAL...18..153S,Smoot:1992td,Kovac:2002fg,Readhead:2004xg,Bennett:2012zja,Crites:2014prc,Naess:2014wtr,Ade:2013hjl,Ade:2015tva}. Current and future measurements \cite{Adam:2015rua,2011JCAP...07..025K,2009arXiv0906.1188B,2014SPIE.9153E..1LL,Andre:2013afa,Andre:2013nfa} will be able to investigate more subtle effects, such as the kinetic Sunyaev-Zel'dovich (kSZ) \cite{1972CoASP...4..173S,Carlstrom:2002na} and CMB spectral distortions \cite{Chluba:2011hw,Khatri:2012tw}.

While modern cosmology has advanced significantly using our understanding from linear perturbation theory, nonlinear contributions become significant at late times and at smaller scales. In the quasi-linear regime, many relevant cosmological observables are usefully described using perturbation theory at higher order. Significant effort has been devoted to understanding structure formation via a range of perturbative techniques (\eg \cite{Bernardeau:2001qr,Sugiyama:2013mpa,Crocce:2005xy,McDonald:2006hf,McDonald:2014dxa,2011JCAP...10..037A,Baumann:2010tm,Carrasco:2012cv,Pajer:2013jj,Hertzberg:2012qn,Blas:2015qsi}). In this work, we consider integrals in standard perturbation theory (SPT), although the methods and code we develop have a broader range of applications.

The next-to-leading-order (``1-loop'') corrections in these perturbative expansions are typically expressed as two-dimensional mode-coupling convolution integrals, which are generically time consuming to evaluate numerically. Recent algorithmic developments have dramatically sped up these computations for {\em scalar} quantities -- those with no dependence on the direction of the observer, such as the matter density or real-space galaxy density. The new algorithms \cite{McEwen:2016fjn,Schmittfull:2016jsw} take advantage of the locality of evolution in perturbation theory, the scale invariance of cold dark matter (CDM) structure formation, and the Fast Fourier Transform (FFT); and work is underway to apply them to 2-loop power spectra as well \cite{Schmittfull:2016yqx}. In a previous paper, we introduced the {\py FAST-PT} implementation of these methods in Python \cite{McEwen:2016fjn}.

However, there are many interesting 1-loop convolution integrals for {\em tensor} quantities -- those with explicit dependence on the observer line of sight, such as those arising for redshift-space distortions. In this case, we need convolution integrals with ``tensor'' kernels:\footnote{The kernel $K$ can be expressed as  a sum of polynomials in the relevant dot products. ``Tensor'' refers to the general transformation properties of the cosmological quantities being considered under a symmetry operation -- in this case, rotations in SO(3). For instance, the momentum density is a rank 1 tensor (a vector) while the IA field is a rank 2 tensor. The scalar case (rank 0) considered in \cite{McEwen:2016fjn} is thus a specific application of this more general framework.}
\begin{equation}
I(k) = \int\dq{1}K(\hat{\bm{q}}_1\cdot\hat{\bm{q}}_2,\hat{\bm{q}}_1\cdot\hat{\bm{k}},\hat{\bm{q}}_2\cdot\hat{\bm{k}},q_1,q_2)P(q_1)P(q_2)~,
\label{eq:tensor_int}
\end{equation}
where $K(\hat{\bm{q}}_1\cdot\hat{\bm{q}}_2,\hat{\bm{q}}_1\cdot\hat{\bm{k}},\hat{\bm{q}}_2\cdot\hat{\bm{k}},q_1,q_2)$ is a tensor mode-coupling kernel, $\bm{k}=\bm{q}_1+\bm{q}_2$, $k=\vert\bm{k}\vert$, and $P(q)$ is the input signal -- typically the linear matter power spectrum -- logarithmically sampled in $q$. Due to the dependence on the direction of $\bm{k}$, the decomposition of these kernels is more complicated than in the scalar case. In this work, we generalize our {\py FAST-PT} algorithm to evaluate these tensor convolution integrals, achieving $\mathcal{O}(N\log N)$ performance as in the scalar case.

This paper is organized as follows: in \S\ref{sec:method} we provide the mathematical basis for our method (\S\ref{subsec:theory}), introduce our algorithm (\S\ref{subsec:algorithm}), and discuss divergences that may arise and how they are resolved (\S\ref{subsec:div}). In section \S\ref{sec:example} we apply our method to several examples: the quadratic intrinsic alignment model (\S\ref{subsec:quadratic_IA}); the Ostriker-Vishniac effect (\S\ref{subsec:OV}); the kinetic polarization of CMB (\S\ref{subsec:kP}); and the 1-loop redshift-space power spectrum (\S\ref{subsec:RSD}). Section \S\ref{sec:summary} summarizes the results. An appendix contains derivations of the relevant mathematical identities. The Python code implementing this algorithm and the examples presented in this paper is publicly available at \url{https://github.com/JoeMcEwen/FAST-PT}.

\section{Method}
\label{sec:method}

In this section we extend the {\py FAST-PT} framework to include the computation of convolution integrals with tensor kernels in the form of Eq.~(\ref{eq:tensor_int})

Our approach is similar to the scalar version of {\py FAST-PT}. We first expand the kernel into several Legendre polynomial products -- the explicit dependence on the direction $\hat{\bm{k}}$ requires an expansion in three angles rather than one (as shown in Eq. \ref{eq:general_kernel} and \ref{eq:legendre}). Second, products of Legendre polynomials are written in spherical harmonics using the addition theorem, where the required combinations of spherical harmonics are constrained by Wigner $3j$ symbols and preserve angular momentum (as in Eq. \ref{eqn:legendre_to_spherical}). Third, in configuration space, the integral of each term in the expansion can be further transformed into a product of several one-dimensional integrals (as in Eq. \ref{eq:J-r} and \ref{eq:introB}), which can be quickly performed by assuming a (biased) log-periodic power spectrum and employing FFTs (as in Eq. \ref{eq:Pk_discrete} and \ref{eq:discrete_J-k}).

We will first provide the theory in \S\ref{subsec:theory} and then briefly introduce our algorithm in \S\ref{subsec:algorithm}. Finally, in \S\ref{subsec:div} we will discuss physical divergence problems that can arise and the way to solve them through the choice of appropriate biasing of the log-periodic power spectrum.

\subsection{Transformation To 1D Integrals}\label{subsec:theory}

In general, the kernel function $K$ can be decomposed as a summation of terms
\begin{align}
K(\hat{\bm{q}}_1\cdot\hat{\bm{q}}_2,\hat{\bm{q}}_1\cdot\hat{\bm{k}},\hat{\bm{q}}_2\cdot\hat{\bm{k}},q_1,q_2) = \sum_{\ell_1,\ell_2,\ell,\alpha,\beta}A_{\ell_1\ell_2\ell}^{\alpha\beta}\mathcal{P}_{\ell}(\hat{\bm{q}}_1\cdot\hat{\bm{q}}_2)\mathcal{P}_{\ell_1}(\hat{\bm{k}}\cdot\hat{\bm{q}}_2)\mathcal{P}_{\ell_2}(\hat{\bm{k}}\cdot\hat{\bm{q}}_1)q_1^\alpha q_2^\beta~,
\label{eq:general_kernel}
\end{align}
where $\mathcal{P}_{\ell}$ are the Legendre polynomials, and the $A_{\ell_1\ell_2\ell}^{\alpha\beta}$ coefficients specify the components of a particular kernel. For general angular dependences the sum may require an infinite number of terms. However the kernels that appear in CDM perturbation theory and galaxy biasing theory are composed of a finite number of terms in a polynomial expansion. This decomposition leads us to consider integrals of the form
\begin{equation}
f(k) = \int\dq{1}\mathcal{P}_{\ell}(\hat{\bm{q}}_1\cdot\hat{\bm{q}}_2)\mathcal{P}_{\ell_1}(\hat{\bm{k}}\cdot\hat{\bm{q}}_2)\mathcal{P}_{\ell_2}(\hat{\bm{k}}\cdot\hat{\bm{q}}_1)q_1^\alpha q_2^\beta P(q_1)P(q_2)~.
\label{eq:legendre}
\end{equation}

The product of Legendre polynomials can be decomposed into spherical harmonics by the addition theorem. Using the result presented in Appendix~\ref{app:deri:P_to_Y}, we can write the product of three Legendre polynomials in terms of spherical harmonics and Wigner $3j$ symbols:
\begin{align}
&\mathcal{P}_{\ell}(\hat{\bm{q}}_1\cdot\hat{\bm{q}}_2)\mathcal{P}_{\ell_2}(\hat{\bm{q}}_1\cdot\hat{\bm{k}})\mathcal{P}_{\ell_1}(\hat{\bm{q}}_2\cdot\hat{\bm{k}})\nonumber\\
&=\sum_{J_1,J_2,J_k}C_{\ell_1\ell_2\ell}^{J_1J_2J_k}\sum_{M_1,M_2,M_k}Y_{\scriptscriptstyle J_1M_1}(\hat{\bm{q}}_1)Y_{\scriptscriptstyle J_2M_2}(\hat{\bm{q}}_2)Y_{\scriptscriptstyle J_kM_k}(\hat{\bm{k}})\left(\begin{array}{ccc}
J_1 & J_2 & J_k\\ M_1 & M_2 & M_k
\end{array}\right)~,
\label{eqn:legendre_to_spherical}
\end{align}
with coefficients given by
\begin{align}
C_{\ell_1\ell_2\ell}^{J_1J_2J_k}=&(4\pi)^{3/2}(-1)^{\ell_1+\ell_2+{\ell}}\nonumber\\
 &\times\sqrt{(2J_1+1)(2J_2+1)(2J_k+1)}\left(\begin{array}{ccc}
J_1 & \ell_2 & \ell\\ 0& 0&0
\end{array}\right)
\left(\begin{array}{ccc}
\ell_1 & J_2 & \ell\\ 0& 0&0
\end{array}\right)
\left(\begin{array}{ccc}
\ell_1 & \ell_2 & J_k\\ 0& 0&0
\end{array}\right)
\left\lbrace\begin{array}{ccc}
J_1 & J_2 & J_k\\ \ell_1 & \ell_2 & \ell
\end{array}\right\rbrace~,
\end{align}
where we have used the $3j$ and $6j$ symbols, denoted by ( ) and \{ \}, respectively. The integers $M_1,M_2,M_k$ satisfy the selection rule
$M_1+M_2+M_k=0$.
The coefficients $C_{\ell_1\ell_2\ell}^{J_1J_2J_k}$ map the product of spherical harmonics in Eq.~(\ref{eqn:legendre_to_spherical}), written in terms of the $J_1,J_2,J_k$ basis, to the original $\ell_1,\ell_2, \ell$ basis of Legendre polynomials. Upon replacing the product of Legendre polynomials in Eq.~(\ref{eq:legendre}) with Eq.~(\ref{eqn:legendre_to_spherical}) (omitting the coefficients $C_{\ell_1\ell_2\ell}^{J_1J_2J_k}$), we arrive at an integral over the product of three spherical harmonics, which we will denote as $I_{\scriptscriptstyle J_1J_2J_k}^{\alpha\beta}(k)$. For each combination of $J_1,J_2,J_k$, we have
\begin{align}
I_{\scriptscriptstyle J_1J_2J_k}^{\alpha\beta}(k)&=\sum_{M_1M_2M_k}\int\dq{1}P(q_1)P(q_2)Y_{\scriptscriptstyle J_1M_1}(\hat{\bm{q}}_1)Y_{\scriptscriptstyle J_2M_2}(\hat{\bm{q}}_2)Y_{\scriptscriptstyle J_kM_k}(\hat{\bm{k}})q_1^\alpha q_2^\beta\left(\begin{array}{ccc}
J_1& J_2 & J_k \\
M_1 & M_2 & M_k
\end{array}\right)\nonumber\\
&\equiv\sum_{M_k}Y_{\scriptscriptstyle J_kM_k}(\hat{\bm{k}})T_{\scriptscriptstyle J_1J_2J_kM_k}^{\alpha\beta}(\bm{k})~,
\end{align}
where we have defined
\begin{align}
T_{\scriptscriptstyle J_1J_2J_kM_k}^{\alpha\beta}(\bm{k})&\equiv\sum_{M_1M_2}\left(\begin{array}{ccc}
J_1& J_2 & J_k \\
M_1 & M_2 & M_k
\end{array}\right)H_{\scriptscriptstyle J_1M_1J_2M_2}^{\alpha\beta}(\bm{k})~~{\rm and}\\
H_{\scriptscriptstyle J_1M_1J_2M_2}^{\alpha\beta}(\bm{k})&\equiv\int\dq{1}P(q_1)P(q_2)Y_{\scriptscriptstyle J_1M_1}(\hat{\bm{q}}_1)Y_{\scriptscriptstyle J_2M_2}(\hat{\bm{q}}_2)q_1^\alpha q_2^\beta~.
\end{align}
We can separate $H_{J_1M_1J_2M_2}^{\alpha\beta}(\bm{k})$ into a product of two integrals, respectively over $\bm{q}_1$ and $\bm{q}_2$, by Fourier transforming to configuration space
\begin{align}
H_{\scriptscriptstyle J_1M_1J_2M_2}^{\alpha\beta}(\bm{r})=&\int\dq{1}\dq{2}e^{i(\bm{q}_1+\bm{q}_2)\cdot\bm{r}}q_1^\alpha q_2^\beta P(q_1)P(q_2)Y_{\scriptscriptstyle J_1M_1}(\hat{\bm{q}}_1)Y_{\scriptscriptstyle J_2M_2}(\hat{\bm{q}}_2)\nonumber\\
=&\bar{H}_{\scriptscriptstyle J_1J_2}^{\alpha\beta}(r)Y_{\scriptscriptstyle J_1M_1}(\hat{\bm{r}})Y_{\scriptscriptstyle J_2M_2}(\hat{\bm{r}})~,
\label{eq:H}
\end{align}
where we have used the plane wave expansion (Eq.~\ref{eq:planewave}) together with orthogonality relations (Eq.~\ref{eq:sph_ortho}) to arrive at the equality. We have also defined
\begin{equation}
\bar{H}_{\scriptscriptstyle J_1J_2}^{\alpha\beta}(r)\equiv\frac{(4\pi)^2 i^{\scriptscriptstyle J_1+J_2}}{(2\pi)^6}\int_0^\infty dq_1\ q_1^{2+\alpha}P(q_1)j_{\scriptscriptstyle J_1}(q_1r)\int_0^\infty dq_2\ q_2^{2+\beta}P(q_2)j_{\scriptscriptstyle J_2}(q_2r)~,
\label{eq:Hbar}
\end{equation}
where $j_{\scriptscriptstyle J}(qr)$ are the spherical Bessel functions. Substituting Eq.~(\ref{eq:H}) into the definition of $T_{\scriptscriptstyle J_1J_2J_kM_k}^{\alpha\beta}$ we obtain
\begin{align}
T_{\scriptscriptstyle J_1J_2J_kM_k}^{\alpha\beta}(\bm{r})&=\sum_{M_1M_2}\left(\begin{array}{ccc}
J_1& J_2 & J_k \\
M_1 & M_2 & M_k
\end{array}\right)H_{\scriptscriptstyle J_1M_1J_2M_2}^{\alpha\beta}(\bm{r})\nonumber\\
&=\bar{H}_{\scriptscriptstyle J_1J_2}^{\alpha\beta}(r)\sum_{M_1M_2}\left(\begin{array}{ccc}
J_1& J_2 & J_k \\
M_1 & M_2 & M_k
\end{array}\right)Y_{\scriptscriptstyle J_1M_1}(\hat{\bm{r}})Y_{\scriptscriptstyle J_2M_2}(\hat{\bm{r}})\nonumber\\
&=\bar{H}_{\scriptscriptstyle J_1J_2}^{\alpha\beta}(r)a_{\scriptscriptstyle J_1J_2J_k}Y_{\scriptscriptstyle J_kM_k}^*(\hat{\bm{r}})~,
\label{eq:T-r}
\end{align}
where
\begin{equation}
a_{\scriptscriptstyle J_1J_2J_k}\equiv\sqrt{\frac{(2J_1+1)(2J_2+1)}{4\pi(2J_k+1)}}\left(\begin{array}{ccc}
J_1 & J_2 & J_k\\
0 & 0 & 0
\end{array}\right)~.
\label{eq:a_JJJ}
\end{equation}
The derivation of Eqs.~(\ref{eq:T-r}) and (\ref{eq:a_JJJ}) is provided in Appendix (\ref{app:deri:T-r}). Fourier transforming back to $k$-space, we obtain
\begin{align}
T_{\scriptscriptstyle J_1J_2J_kM_k}^{\alpha\beta}(\bm{k})=&\int d^3r T_{\scriptscriptstyle J_1J_2J_kM_k}^{\alpha\beta}(\bm{r})e^{-i\bm{k}\cdot\bm{r}}\nonumber\\
=&a_{\scriptscriptstyle J_1J_2J_k}\int r^2dr\bar{H}_{\scriptscriptstyle J_1J_2}^{\alpha\beta}(r)\int d^2\hat{\bm{r}} Y_{\scriptscriptstyle J_kM_k}^*(\hat{\bm{r}})e^{-i\bm{k}\cdot\bm{r}}\nonumber\\
=&a_{\scriptscriptstyle J_1J_2J_k}\int r^2dr\bar{H}_{\scriptscriptstyle J_1J_2}^{\alpha\beta}(r)\int d^2\hat{\bm{r}} Y_{\scriptscriptstyle J_kM_k}^*(\hat{\bm{r}})4\pi\sum_{\ell'm'}(-i)^{\ell'}j_{\ell'}(kr)Y_{\ell'm'}^*(\hat{\bm{k}})Y_{\ell'm'}(\hat{\bm{r}})\nonumber\\
=&a_{\scriptscriptstyle J_1J_2J_k}\int r^2dr\bar{H}_{\scriptscriptstyle J_1J_2}^{\alpha\beta}(r) 4\pi\sum_{\ell'm'}(-i)^{\ell'}j_{\ell'}(kr)Y_{\ell'm'}^*(\hat{\bm{k}})\delta_{\scriptscriptstyle \ell'J_k}\delta_{\scriptscriptstyle m'M_k}\nonumber\\
=&4\pi (-i)^{\scriptscriptstyle J_k}a_{\scriptscriptstyle J_1J_2J_k}\int r^2dr\bar{H}_{\scriptscriptstyle J_1J_2}^{\alpha\beta}(r)j_{\scriptscriptstyle J_k}(kr)Y_{\scriptscriptstyle J_kM_k}^*(\hat{\bm{k}})~,
\label{eq:T-k}
\end{align}
where in the third equality we have used the plane wave expansion (Eq.~\ref{eq:planewave}), and in the fourth equality used the orthogonality relation between spherical harmonics (Eq.~\ref{eq:sph_ortho}). Combining the results from Eq.~(\ref{eq:Hbar}), (\ref{eq:T-k}), (\ref{eq:a_JJJ}), we arrive at
\begin{align}
I_{\scriptscriptstyle J_1J_2J_k}^{\alpha\beta}(k)&=4\pi (-i)^{\scriptscriptstyle J_k}a_{\scriptscriptstyle J_1J_2J_k}\int r^2dr\bar{H}_{\scriptscriptstyle J_1J_2}^{\alpha\beta}(r)j_{\scriptscriptstyle J_k}(kr)\sum_{M_k} Y_{\scriptscriptstyle J_kM_k}(\hat{\bm{k}})Y_{\scriptscriptstyle J_kM_k}^*(\hat{\bm{k}})\nonumber\\
&=(-i)^{\scriptscriptstyle J_k}(2J_k+1)a_{\scriptscriptstyle J_1J_2J_k}\int r^2dr\bar{H}_{\scriptscriptstyle J_1J_2}^{\alpha\beta}(r)j_{\scriptscriptstyle J_k}(kr)\nonumber\\
&=(-1)^{J_k+(J_1+J_2+J_k)/2}\sqrt{\frac{(2J_1+1)(2J_2+1)(2J_k+1)}{64\pi^9}}\left(\begin{array}{ccc}
J_1 &J_2 & J_k\\
0&0&0
\end{array}\right)
\nonumber \\ & ~~~~\times
\int r^2dr J_{\scriptscriptstyle J_1J_2}^{\alpha\beta}(r)j_{\scriptscriptstyle J_k}(kr)~,
\label{eq:I-k}
\end{align}
where $J_1+J_2+J_k$ must be even for the $3j$ symbol to be non-zero, and $J_{\scriptscriptstyle J_1J_2}^{\alpha\beta}(r)$ is defined by
\begin{equation}
J_{\scriptscriptstyle J_1J_2}^{\alpha\beta}(r)\equiv\left[\int_0^\infty dq_1\ q_1^{2+\alpha}P(q_1)j_{\scriptscriptstyle J_1}(q_1r)\right]\left[\int_0^\infty dq_2\ q_2^{2+\beta}P(q_2)j_{\scriptscriptstyle J_2}(q_2r)\right]~.
\label{eq:J-r}
\end{equation}
Combining Eq. (\ref{eq:I-k}) and (\ref{eqn:legendre_to_spherical}) we can rewrite the integral (\ref{eq:legendre}) as
\begin{align}
&\int\dq{1}\mathcal{P}_{\ell}(\hat{\bm{q}}_1\cdot\hat{\bm{q}}_2)\mathcal{P}_{\ell_1}(\hat{\bm{k}}\cdot\hat{\bm{q}}_2)\mathcal{P}_{\ell_2}(\hat{\bm{k}}\cdot\hat{\bm{q}}_1)q_1^\alpha q_2^\beta P(q_1)P(q_2)\nonumber\\
&=\sum_{J_1,J_2,J_k}C_{\ell_1\ell_2\ell}^{J_1J_2J_k}I_{\scriptscriptstyle J_1J_2J_k}^{\alpha\beta}(k)=\sum_{J_1,J_2,J_k}B_{\ell_1\ell_2\ell}^{J_1J_2J_k}\int r^2dr J_{\scriptscriptstyle J_1J_2}^{\alpha\beta}(r)j_{\scriptscriptstyle J_k}(kr)~,
\label{eq:introB}
\end{align}
where the coefficients $B_{\ell_1\ell_2\ell}^{J_1J_2J_k}$ are given by
\begin{align}
B_{\ell_1\ell_2\ell}^{J_1J_2J_k}\equiv&C_{\ell_1\ell_2\ell}^{J_1J_2J_k}(-1)^{J_k+(J_1+J_2+J_k)/2}\sqrt{\frac{(2J_1+1)(2J_2+1)(2J_k+1)}{64\pi^9}}\left(\begin{array}{ccc}
J_1 &J_2 & J_k\\
0&0&0
\end{array}\right)\nonumber\\
=&(-1)^{\ell+\frac{J_1+J_2+J_k}{2}}\times\frac{(2J_1+1)(2J_2+1)(2J_k+1)}{\pi^3}\nonumber\\
&\times\left(\begin{array}{ccc}
J_1 & \ell_2 & \ell\\ 0& 0&0
\end{array}\right)
\left(\begin{array}{ccc}
\ell_1 & J_2 & \ell\\ 0& 0&0
\end{array}\right)
\left(\begin{array}{ccc}
\ell_1 & \ell_2 & J_k\\ 0& 0&0
\end{array}\right)
\left(\begin{array}{ccc}
J_1 & J_2 & J_k\\ 0& 0&0
\end{array}\right)
\left\lbrace\begin{array}{ccc}
J_1 & J_2 & J_k\\ \ell_1 & \ell_2 & \ell
\end{array}\right\rbrace~.
\label{eq:coeff_B}
\end{align}
The evaluation of $J_{\scriptscriptstyle J_1J_2}^{\alpha\beta}(r)$ is similar to the analogous quantity in scalar {\py FAST-PT}. For notational simplicity, we define the last integral in Eq.~(\ref{eq:introB}) as
\begin{align}
\label{eq:J_fancy}
\mathcal{J}^{\alpha \beta}_{\scriptscriptstyle J_1 J_2 J_k}(k) = \int r^2 dr J^{\alpha \beta}_{\scriptscriptstyle J_1 J_2}(r) j_{\scriptscriptstyle J_k}(kr) ~. 
\end{align} 
Eq.~(\ref{eq:J_fancy}) is similar in structure to Eq.~(2.19) of \cite{McEwen:2016fjn}. As such, we can easily generalize the {\py FAST-PT} framework to evaluate integrals in the form of Eq.~(\ref{eq:J_fancy}). 

Note that some (scalar) 2-loop integrals have similar structure to the tensor 1-loop integrals considered here. In recent work, Ref.~\cite{Schmittfull:2016yqx} employed similar techniques involving Wigner $6j$ symbols to deal with these 2-loop integrals, although the implementations are somewhat different.

\subsection{Algorithm}\label{subsec:algorithm}
\subsubsection{Implementation For $\mathcal{J}_{\scriptscriptstyle J_1J_2J_k}^{\alpha\beta}(k)$ Integral}\label{subsub-implement}
We adopt the discrete Fourier transformation of the power spectrum as discussed in the first {\py FAST-PT} paper \cite{McEwen:2016fjn},
\begin{align}
c_m=\displaystyle W_m \sum_{q=0}^{N-1}  \frac{P(k_q)}{k_q^{\nu_1}} e^{- 2 \pi im q/N}
~~~\rightarrow~~~
P_{\rm filtered}(k_q)=\displaystyle \sum_{m=-N/2}^{N/2} c_m k_q^{\nu_1 + i \eta_m}~, 
\label{eq:Pk_discrete}
\end{align}
where $N$ is the size of the input power spectrm, $\eta_m= m \times 2\pi/(N\Delta) $, $m=-N/2, -N/2 + 1,...,N/2-1, N/2$, $\nu_1$ is the bias index, and $\Delta$ is the linear spacing, \ie $k_q=k_0\exp(q \Delta)$ with $k_0$ being the smallest value in the $k$ array. Similarly, $c'_n$ are the Fourier coefficients of the power spectrum with bias index $\nu_2$. The physics of the bias has been discussed in \cite{McEwen:2016fjn}\footnote{The bias is introduced to solve the numerical divergences arising from the Fourier transform. By performing the Fourier transform, we assume the input power spectrum to be periodic, so that there are infinite ``satellite'' power spectra on both low and high $k$ sides. To avoid infinite contribution from the satellites, appropriate bias values are required.} and the choice of its value will be discussed in \S\ref{subsub:divbias}. For a real power spectrum the Fourier coefficients obey $c_m^\ast=c_{-m},~c_n'^\ast=c_{-n}'$. $W_m$ is a window function\footnote{The window function we use is a smoothing function described in Appendix C of \cite{McEwen:2016fjn}.} used to smooth the edges of the Fourier coefficient array of the biased power spectrum (\eg from the cutoffs in $k$), hence smoothing over the noise and sharp features in the power spectrum, as well as prevent them from propagating non-locally in the ``filtered'' power spectrum. The ``filtered'' power spectrum is then treated as the input power spectrum and its $c_m$'s are used for calculations afterwards. Following Eq. (2.17) in \cite{McEwen:2016fjn}, we can write Eq. (\ref{eq:J-r}) as\footnote{The major step is substituting the expansions of the power spectra into Eq.~(\ref{eq:J-r}), and utilizing the formula: $\int_0^\infty dt\ t^\kappa J_{\mu}(t)=2^\kappa g(\mu,\kappa)$ for $\Re\kappa<1/2,\ \Re(\kappa+\mu)>-1$, where the Bessel function of the first kind $J_\mu$ is related to the spherical Bessel function by $J_\mu(t)=\sqrt{2t/\pi}j_{\mu-1/2}(t)$, and $g(\mu,\kappa)$ is defined in Eq.~(\ref{eq:g-mu-kappa}).}
\begin{equation}
J_{\scriptscriptstyle J_1J_2}^{\alpha\beta}(r)=\frac{\pi}{2}\sum_{m=-N/2}^{N/2} \sum_{n=-N/2}^{N/2}c_m c_n' g_{\alpha m}g_{\beta n}2^{Q_{\alpha m}+Q_{\beta n}} r^{-6-\nu_1-\nu_2-\alpha-\beta-i\eta_m-i\eta_n}~,
\label{eq:discrete_J-r}
\end{equation}
where $g_{\alpha m}\equiv g(J_1+\frac{1}{2},Q_{\alpha m})$, $g_{\beta n}\equiv g(J_2+\frac{1}{2},Q_{\beta n})$, $Q_{\alpha m}\equiv \frac{3}{2}+\nu_1+\alpha+i\eta_m$, $Q_{\beta n}\equiv \frac{3}{2}+\nu_2+\beta+i\eta_n$, and
\begin{equation}
g(\mu,\kappa)\equiv\frac{\Gamma[(\mu+\kappa+1)/2]}{\Gamma[(\mu-\kappa+1)/2]}~.
\label{eq:g-mu-kappa}
\end{equation}
The integral then becomes
\begin{align}
&\mathcal{J}_{\scriptscriptstyle J_1J_2J_k}^{\alpha\beta}(k_q)\equiv \int_0^\infty dr\ r^2 J_{\scriptscriptstyle J_1J_2}^{\alpha\beta}(r) j_{\scriptscriptstyle J_k}(k_q r)\nonumber\\
=&\frac{\pi}{2}\sum_{m=-N/2}^{N/2} \sum_{n=-N/2}^{N/2}c_m g_{\alpha m} c_n' g_{\beta n}2^{Q_{\alpha m}+Q_{\beta n}}\int_0^\infty dr\ j_{\scriptscriptstyle J_k}(k_q r)r^{-4-\nu_1-\nu_2-\alpha-\beta-i\eta_m-i\eta_n}\nonumber\\
=&\frac{\pi}{2}\sum_{m=-N/2}^{N/2} \sum_{n=-N/2}^{N/2}c_m g_{\alpha m} c_n' g_{\beta n}2^{Q_{\alpha m}+Q_{\beta n}}k_q^{Q_{\alpha m}+Q_{\beta n}}
\int_0^\infty dr\ j_{\scriptscriptstyle J_k}(r)r^{-4-\nu_1-\nu_2-\alpha-\beta-i\eta_m-i\eta_n}\nonumber\\
=&\left(\frac{\pi}{2}\right)^{\frac{3}{2}}\sum_{m=-N/2}^{N/2} \sum_{n=-N/2}^{N/2}c_m g_{\alpha m} c_n' g_{\beta n}2^{Q_{\alpha m}+Q_{\beta n}}k_q^{Q_{\alpha m}+Q_{\beta n}}\nonumber\\
&\times\int_0^\infty dr\ J_{\scriptscriptstyle J_k+\frac{1}{2}}(r)r^{-\frac{9}{2}-\nu_1-\nu_2-\alpha-\beta-i\eta_m-i\eta_n}\nonumber\\
=&\left(\frac{\pi}{2}\right)^{\frac{3}{2}}\sum_{m=-N/2}^{N/2} \sum_{n=-N/2}^{N/2}c_m g_{\alpha m} c_n' g_{\beta n}2^{Q_{\alpha m}+Q_{\beta n}}k_q^{Q_{\alpha m}+Q_{\beta n}}\nonumber\\
&\times 2^{-\frac{9}{2}-\nu_1-\nu_2-\alpha-\beta-i\eta_m-i\eta_n}g\left(J_k+\frac{1}{2},-\frac{9}{2}-\nu_1-\nu_2-\alpha-\beta-i\eta_m-i\eta_n\right)\nonumber\\
=&\frac{\pi^{3/2}}{8}\sum_{m=-N/2}^{N/2} \sum_{n=-N/2}^{N/2}c_m g_{\alpha m} c_n' g_{\beta n}k_q^{Q_{\alpha m}+Q_{\beta n}}
g\left(J_k+\frac{1}{2},-\frac{9}{2}-\nu_1-\nu_2-\alpha-\beta-i\eta_m-i\eta_n\right)~.
\end{align}
We define $\tau_h\equiv\eta_m+\eta_n$ and $Q_h\equiv Q_{\alpha m}+Q_{\beta n}$, which only depends on the sum $m+n$. We write the double summation over $m$ and $n$ as a discrete convolution, indexed by $h$, such that $h=m+n=\lbrace-N,-N+1,\cdots,N-1,N\rbrace$. This leads to
\begin{align}
\mathcal{J}_{\scriptscriptstyle J_1J_2J_k}^{\alpha\beta}(k_q)&=\frac{\pi^{3/2}}{8}\sum_{m=-N/2}^{N/2} \sum_{n=-N/2}^{N/2}c_m g_{\alpha m} c_n' g_{\beta n}k_q^{Q_{h}}
g\left(J_k+\frac{1}{2},-Q_h-\frac{3}{2}\right)\nonumber\\
&=\frac{\pi^{3/2}}{8}\sum_h [c_m g_{\alpha m}\otimes c_n' g_{\beta n}]_h k_q^{Q_{h}}
g\left(J_k+\frac{1}{2},-Q_h-\frac{3}{2}\right)\nonumber\\
&=\frac{\pi^{3/2}}{8}k_q^{3+\nu_1+\nu_2+\alpha+\beta}\sum_h C_h \exp(i\tau_h\ln k_0)\exp(i\tau_h q\Delta) g\left(J_k+\frac{1}{2},-Q_h-\frac{3}{2}\right)\nonumber\\
&=\frac{\pi^{3/2}}{8}k_q^{3+\nu_1+\nu_2+\alpha+\beta} {\rm IFFT}\left[C_h\ g\left(J_k+\frac{1}{2},-Q_h-\frac{3}{2}\right)\right]~,
\label{eq:discrete_J-k}
\end{align}
where $C_h$ is defined as the convolution in the second equality, and IFFT is the discrete inverse Fast Fourier Transform. This derivation is similar to Eq. (2.21) in \cite{McEwen:2016fjn}.

In the algorithm, for each set of $(J_1,J_2,J_k)$ there are 3 FFT operations and 1 convoluton. In our public code, we use the {\py scipy.signal.fftconvolve} routine \cite{scipy} to perform the convolution, which uses the convolution theorem, resulting in 3 additional FFT operations. Thus, for each set of $(J_1,J_2,J_k)$ there are 6 FFT operations executed in total.

\subsubsection{Summary of the Algorithm}\label{quick_review}
From Eq~(\ref{eq:introB}), the tensor convolution integral (\ref{eq:tensor_int}) can be decomposed as
\begin{equation}
I(k) = \sum_{\ell_1,\ell_2,\ell,\alpha,\beta}A_{\ell_1\ell_2\ell}^{\alpha\beta}\sum_{J_1,J_2,J_k}B_{\ell_1\ell_2\ell}^{J_1J_2J_k}\int r^2dr J_{\scriptscriptstyle J_1J_2}^{\alpha\beta}(r)j_{\scriptscriptstyle J_k}(kr)~.
\end{equation}
Our algorithm is thus as follows:
\begin{enumerate}
 \item Given an integral in the form of Eq. (\ref{eq:tensor_int}), expand it in terms of Eq. (\ref{eq:legendre}) to obtain all the non-zero coefficients $A_{\ell_1\ell_2\ell}^{\alpha\beta}~$;
 \item For each combination of $\ell_1,\ell_2,\ell$, use Eq. (\ref{eq:coeff_B}) to calculate all the possible combinations of $J_1,J_2,J_k$ and their corresponding (non-zero) coefficients $B_{\ell_1\ell_2\ell}^{J_1J_2J_k}$;
\item For all the possible combinations of $J_1,J_2,J_k$, calculate $J_{\scriptscriptstyle J_1J_2}^{\alpha\beta}(r)$ and perform the Hankel transform integration (see \S\ref{subsub-implement} for the detailed implementation);
\item Sum up all the terms to obtain the result.
\end{enumerate}
The criteria for non-zero $B_{\ell_1\ell_2\ell}^{J_1J_2J_k}$ can be obtained from the properties of the Wigner $3j$ symbols. From Eq. (\ref{eq:coeff_B}) we have
\begin{equation}
\vert \ell_1-\ell_2\vert\leq J_k\leq \ell_1+\ell_2~,~~\vert \ell-\ell_2\vert\leq J_1\leq \ell+\ell_2~,~~\vert \ell-\ell_1\vert\leq J_2\leq \ell+\ell_1~,
\end{equation}
\begin{equation}
\vert J_1-J_2\vert\leq J_k\leq J_1+J_2~,
\end{equation}
and
\begin{equation}
J_1+\ell_2+\ell={\rm even}~,~~\ell_1+J_2+\ell={\rm even}~,~~\ell_1+\ell_2+J_k={\rm even}~.
\label{eq:even_condition}
\end{equation}
The condition that ``$J_1+J_2+J_k={\rm even}$'' is redundant since it can be infered from the conditions (Eq.~\ref{eq:even_condition}).\footnote{Summing up the three equations in Eq. (\ref{eq:even_condition}) we have $J_1+J_2+J_k+2(\ell_1+\ell_2+\ell)={\rm even}$, which leads to $J_1+J_2+J_k={\rm even}$.}.

\subsection{Removing Possible Divergences} \label{subsec:div}

Note that the algorithm we have presented in this section is only for the ``$P_{22}(k)$''-type integrals, \ie containing two power spectra $P(q_1)P(q_2)$ in the integrand as in Eq.~(\ref{eq:tensor_int}). In \S\ref{subsub-RSD-conver} we will encounter integrals containing $P(q_1)P(k)$ or $P(q_2)P(k)$, which can be reduced to one-dimensional integrals, analogous to $P_{13}(k)$ in 1-loop SPT (for details on our algorithm of $P_{22}$ and $P_{13}$, see \cite{McEwen:2016fjn}). We first focus on the $P_{22}(k)$-type integrals, where two potential types of divergence may emerge in this algorithm.

\subsubsection{Divergence From Kernel Expansions}\label{subsub:divkernel}

When we expand the kernel into the Legendre polynomial form, the integral (\ref{eq:legendre}) can be divergent for some combinations of $\ell,\ell_1,\ell_2,\alpha,\beta$, even though the sum of all terms will be convergent for physical observables. If the input power spectrum is the linear matter power spectrum $\Pl(k)$, for $q_1 \gg k$, $\bm{q}_2 \approx -\bm{q}_1$, and the power spectra, $\Pl(q_1)$ and $\Pl(q_2)$, both scale as $q_1^{-3}$. Thus the integral (\ref{eq:legendre}) is proportional to $\int dq_1\,q_1^{\alpha+\beta-4}$ for $\ell_1= \ell_2$. Convergence requires that $\alpha+\beta <3$. For $\ell_1\neq \ell_2$, this constraint is relaxed due to suppression from the angular integral.

For $q_1 \ll k$, $\bm{q}_2 \approx \bm{k}$, so that $\Pl(q_1)\propto q_1^{n_s}$ and $\Pl(q_2)\propto k^{n_{\rm eff}(k)}$, where $n_s\sim 1$ is the primordial spectral index of the matter power spectrum, and $n_{\rm eff}(k)$ is the effective spectral index at $k$. The integral is then proportional to $\int dq_1\,q_1^{\alpha+n_s+2}$, leading to the requirement: $\alpha>-3-n_s$ for $\ell=\ell_2$. Similarly, for $q_2$ small, we get $\beta>-3-n_s$ for $\ell=\ell_1$. As before, these constraints are relaxed if $\ell\neq\ell_2$ or $\ell\neq\ell_1$.

Violations of these criteria have to be removed by regularization, specifically canceling the divergent parts. None of the examples in the next section have such a divergence (although see \S\ref{subsub-RSD-conver} for a discussion of a separate numerical divergence which is treated analytically).

\subsubsection{Divergence From Periodic Power Spectrum and Choice of Bias Indices}\label{subsub:divbias}

As discussed in \cite{McEwen:2016fjn}, the use of FFTs enforces a periodic power spectrum which can lead to unphysical divergences for certain choices of the power-law bias. This generalized implementation of {\py FAST-PT} has more freedom in the choice of bias indices $\nu_1,\nu_2$, compared with the original ``scalar'' version. First, it allows the use of two different bias indices $\nu_1,\nu_2$ for the two input power spectra, instead of one fixed $\nu$. Second, it allows the bias indices to change for different Legendre integrals (\ref{eq:legendre}). We now discuss our choice of $\nu_1,\nu_2$.

In {\py FAST-PT}, we expand the input power spectra $\Pl(q_1),\Pl(q_2)$ into sums over power-law spectra $q_1^{\nu_1+i\eta_m}$ and $q_2^{\nu_2+i\eta_n}$. The real parts of the exponents, \ie the bias indices $\nu_1,\nu_2$, will affect the convergence of the integrals.

\begin{figure}
\centering
\includegraphics[scale=0.4]{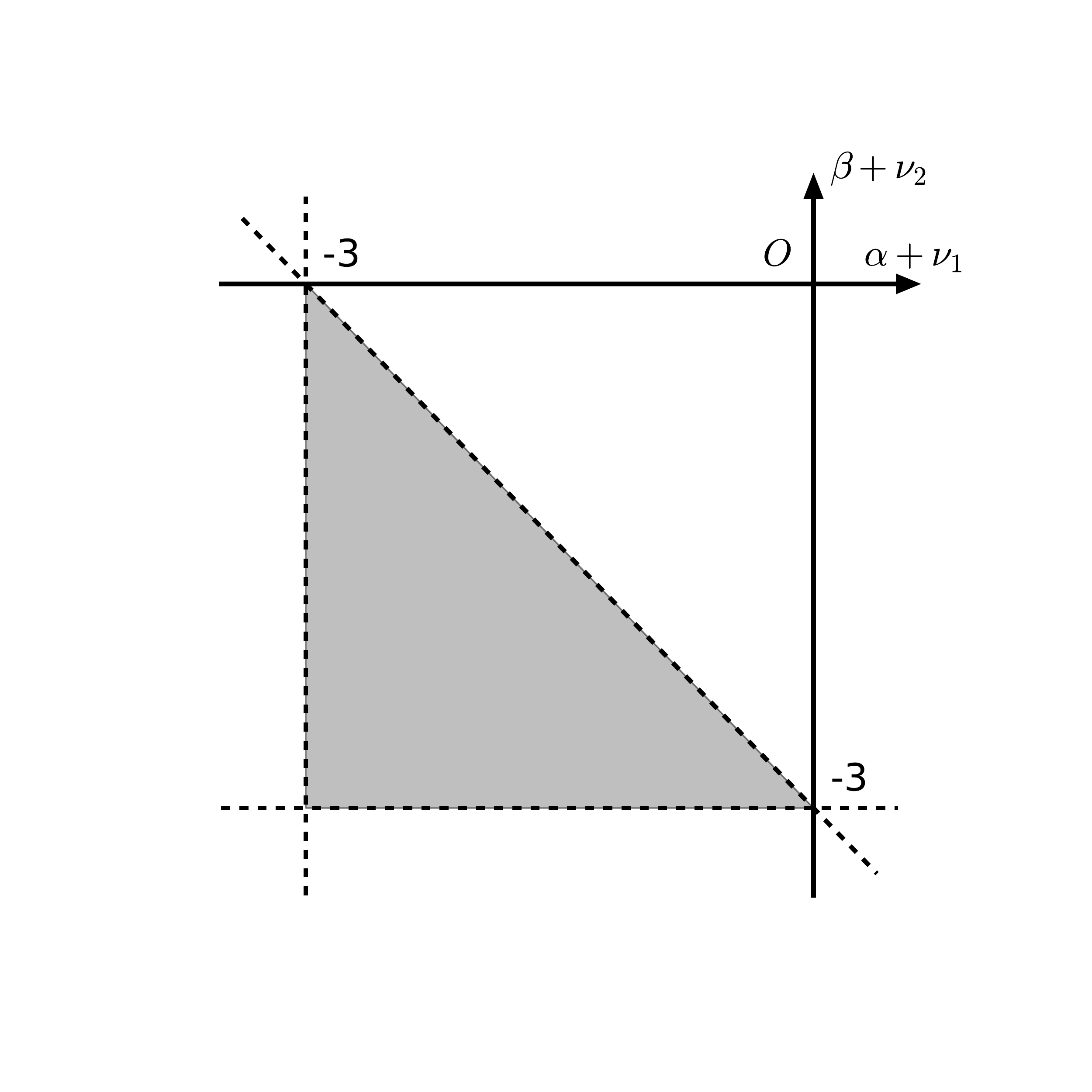}
\caption{The convergence region of the bias indices $\nu_1,\nu_2$ is indicated by the shaded region.}
\label{fig:bias_choice}
\end{figure}

Using a similar argument as in the previous subsection, for large $q_1$, we will have $\Pl(q_1)\propto q_1^{\nu_1},\Pl(q_2)\propto q_1^{\nu_2}$. Working out the integral, we end up with the criterion: $\nu_1+\alpha+\nu_2+\beta<-3$ for $\ell_1=\ell_2$. For small $q_1$, we get $\alpha+\nu_1>-3$ for $\ell=\ell_2$; similarly for small $q_2$, we get $\beta+\nu_2>-3$ for $\ell=\ell_1$. These constraints are relaxed if $\ell\neq\ell_2$ or $\ell\neq\ell_1$. We plot the convergence region in Figure \ref{fig:bias_choice}.

In our code, we take $\nu_1=-2-\alpha$ and $\nu_2=-2-\beta$ for all cases to satisfy the above conditions. Note that the choice of different bias values for different components of a given observable is technically non-physical since the choice of bias specifies the properties of the ``universe in which the calculation is done. However, if the input $k$-range (or zero-padding) is sufficient, this effect is negligible on scales of interest\footnote{In principle, different bias indices could lead to slightly different integral results due to contributions from the periodic ``satellite'' power spectra. However, when the input $k$-range or zero-padding is sufficient, these artificial contributions become negligible. When the bias indices are chosen inside the convergence region in Fig.\ref{fig:bias_choice}, we can always find a sufficient $k$-range, while outside the region, there may be no sufficient range. To test the stability of the results, we compared the OV power spectrum (Eq. \ref{eq:Sk-OV}) obtained using the bias indices $\nu_1=-2-\alpha, \nu_2=-2-\beta$ to the result obtained with the indices $\nu_1=-2.5-\alpha, \nu_2=-2.5-\beta$, and found that the maximum fractional difference over the range 0.003-10 $h/$Mpc is less than 3$\times 10^{-7}$.}. The fixed biasing scheme ($\nu=-2$) employed for scalar quantities in \cite{McEwen:2016fjn} avoids this issue. However, because one component of $P_{22}$ violates $\alpha + \nu > -3$ (for $\ell=0$) under this fixed biasing, we required analytic regularization to enforce Galilean invariance and remove the formally infinite contribution to displacements from $k \rightarrow 0$ modes. Those integrals can be performed using the new scheme without the analytic regularization, although in this case a larger input range in $k$ (or additional zero-padding) is required for numerical convergence.

\section{Applications}\label{sec:example}

In this section we apply the {\py FAST-PT} tensor algorithm to several cosmological applications: the quadratic intrinsic alignment model (\S\ref{subsec:quadratic_IA}); the Ostriker-Vishniac effect (\S\ref{subsec:OV}); the kinetic polarization of CMB (\S\ref{subsec:kP}); and the 1-loop redshift-space distortion power spectrum (\S\ref{subsec:RSD}). In each subsection we first briefly review the theory behind the application before expanding the relevant integral(s) into the form of Eq.~(\ref{eq:legendre}) and comparing the output for each case with the results from conventional (and significantly slower) two-dimensional cubature integration. To demonstrate the performance of the code, we provide this comparison out to high wavenumbers ($k=10~h/$Mpc). We caution that the underlying perturbative models are not applicable to the real Universe beyond the the mildly nonlinear regime ($k\sim {\rm few}\times 10^{-1}~h/$Mpc), even though {\py FAST-PT} can still accurately compute the perturbation theory integrals. We envision these examples both as results in and of themselves, and, more importantly, as reference material for other cosmologists who may want to compute 1-loop power spectra with their own kernels and convert them to {\py FAST-PT} format.

Our input linear power spectrum was generated by CAMB \cite{Lewis:1999bs}, assuming a flat $\Lambda$CDM cosmology corresponding to the Planck 2015 results \cite{Ade:2015xua}. We used Python version 3.5.1, {\sc numpy} 1.10.4, and {\sc scipy} 0.17.0. The public code is also compatible with Python 2.

\subsection{Quadratic Intrinsic Alignments Model}
\label{subsec:quadratic_IA}

\subsubsection{Theory}\label{subsub-IA-theory}

Weak gravitational lensing has become one of the most promising probes of the dark matter distribution \cite{Abbott:2016ktf,Hildebrandt:2016iqg}. The observed shapes of galaxies are weakly distorted (``sheared'') by the gravitational potential of the large-scale structure along the line of sight. Correlations in observed shapes tell us about the projected matter distribution. However, weak lensing suffers from several systematic effects, one of which is intrinsic correlations between galaxy ellipticities, known as ``intrinsic alignments'' (IA) \cite{Troxel:2014dba,Joachimi:2015mma}. In the weak lensing regime, the intrinsic shapes of galaxies dominate the observed shapes (\ie are much larger than the lensing shear contribution). While the dominant uncorrelated component of intrinsic ellipticities does not affect the correlation of shapes beyond adding noise, the component correlating the ellipticity with the underlying tidal field can bias cosmological inference from weak lensing measurements \cite{Krause:2015jqa}. On the other hand, IA can also serve as a probe of the the cosmological density field as well as the astrophysics of galaxies and halos \cite{Chisari:2013dda}.

On large scales, there are two types of physically-motivated intrinsic galaxy alignment models, the tidal (linear) and quadratic alignment models \cite{Hirata:2004gc,Catelan:2000vm}. The tidal alignment model is based on the assumption that large-scale correlations in the intrinsic ellipticity field of triaxial elliptical galaxies are linearly related to fluctuations in the primordial gravitational tidal field in which the galaxy formed.\footnote{Similar results are obtained from assuming that intrinsic shapes are ``instantaneously'' set by the tidal field at the time of observation (see \cite{Blazek:2015lfa} for further discussion).} In quadratic models (often referred to as ``tidal torquing''), the observed ellipticity of spiral galaxies comes from the inclination of the disk with respect to the line of sight, and hence from the direction of its angular momentum. In this scenario, the tidal field from the large-scale structure will both ``spin-up'' the galaxy as well as provide a torque, contributing to the mean intrinsic ellipticity at second order. In general, once nonlinear effects are included, both tidal alignment and tidal torquing models have contributions from mode coupling integrals of the form of Eq.~\ref{eq:tensor_int} \cite{Blazek:2015lfa}. More generally, these models can be viewed as components in an ``effective expansion'' of IA \cite{blazek16_inprep}, analogous to treatments of galaxy biasing \cite{McDonald2009JCAP...08..020M}.

In the quadratic alignment model \cite{Hirata:2004gc}, the intrinsic alignment $E/B$-mode power spectrum $P_{\tilde{\gamma}_I}^{(EE,BB)}(k)$ contains a convolution integral in the form of
\begin{equation}
P_{\rm IA,quad}^{(EE,BB)}(k)=2\int\dq{1}h_{(E,B)}^2(\hat{\bm{q}}_1,\hat{\bm{q}}_2)\Pl(q_1)\Pl(q_2),
\label{eq:Pk-IA}
\end{equation}
where $\bm{k}=\bm{q}_1+\bm{q}_2$ and $h_E$ and $h_B$ are tensor kernels. If we choose the coordinate system such that $\hat{\bm{k}}=\hat{\bm{z}}$ and $\hat{\bm{x}}$ points to the observer, $h_{(E,B)}$ can be expressed as
\begin{align}
h_E(\hat{\bm{q}}_1,\hat{\bm{q}}_2)=&h_{zz}(\hat{\bm{q}}_1,\hat{\bm{q}}_2)-h_{yy}(\hat{\bm{q}}_1,\hat{\bm{q}}_2)\nonumber\\
=&(\hat{\bm{q}}_1\cdot\hat{\bm{q}}_2)(\hat{\bm{q}}_1\cdot\hat{\bm{k}})(\hat{\bm{q}}_2\cdot\hat{\bm{k}})-\frac{1}{3}(\hat{\bm{q}}_1\cdot\hat{\bm{k}})^2-\frac{1}{3}(\hat{\bm{q}}_2\cdot\hat{\bm{k}})^2\nonumber\\
&-(\hat{\bm{q}}_1\cdot\hat{\bm{q}}_2)(\hat{\bm{q}}_1\cdot\hat{\bm{y}})(\hat{\bm{q}}_2\cdot\hat{\bm{y}})+\frac{1}{3}(\hat{\bm{q}}_1\cdot\hat{\bm{y}})^2+\frac{1}{3}(\hat{\bm{q}}_2\cdot\hat{\bm{y}})^2~,\\
h_B(\hat{\bm{q}}_1,\hat{\bm{q}}_2)=&2h_{zy}(\hat{\bm{q}}_1,\hat{\bm{q}}_2)\nonumber\\
=&\left[(\hat{\bm{q}}_1\cdot\hat{\bm{q}}_2)(\hat{\bm{q}}_2\cdot\hat{\bm{k}})-\frac{2}{3}(\hat{\bm{q}}_1\cdot\hat{\bm{k}})\right](\hat{\bm{q}}_1\cdot\hat{\bm{y}}) \nonumber\\
&+\left[(\hat{\bm{q}}_1\cdot\hat{\bm{q}}_2)(\hat{\bm{q}}_1\cdot\hat{\bm{k}})-\frac{2}{3}(\hat{\bm{q}}_2\cdot\hat{\bm{k}}) \right](\hat{\bm{q}}_2\cdot\hat{\bm{y}})
\end{align}
where we can see that $h_{(E,B)}$ have $\bm{\hat{k}}$ dependence. We have made the Limber approximation in assuming that only modes transverse to the line of sight will contribute to observed correlations, hence $\hat{\bm{n}}=\hat{\bm{x}}$. Note that our choice of the coordinate system is different from the conventions in some previous work where $\hat{\bm{z}}$ is chosen to be along the line of sight. Because the integrand has an azimuthal symmetry around $\bm{k}$, independent of the line-of-sight direction, it is more convenient to work in our coordinate system, although the final results do not depend on this choice.

\subsubsection{Conversion to {\py FAST-PT} Format}\label{subsub-IA-conver}
In spherical coordinates, we have $\hat{\bm{q}}_{i}=(\sin\theta_i\cos\phi_i,\sin\theta_i\sin\phi_i,\cos\theta_i)$ for $i=1,2$. Note that $\phi_1=\phi_2-\pi\equiv\phi$ because $\bm{q}_1$ and $\bm{q}_2$ add up to $\bm{k}$ which is on the $z-$axis. We obtain
\begin{align}
\hat{\bm{q}}_1\cdot\hat{\bm{y}}&=\sin\theta_1\sin\phi,\ \ \ \hat{\bm{q}}_2\cdot\hat{\bm{y}}=-\sin\theta_2\sin\phi,\nonumber\\
\hat{\bm{q}}_1\cdot\hat{\bm{q}}_2&=\cos\theta_1\cos\theta_2-\sin\theta_1\sin\theta_2,\nonumber\\
(\hat{\bm{q}}_1\cdot\hat{\bm{y}})(\hat{\bm{q}}_2\cdot\hat{\bm{y}})&=\left(\hat{\bm{q}}_1\cdot\hat{\bm{q}}_2 - (\hat{\bm{q}}_1\cdot\hat{\bm{k}})(\hat{\bm{q}}_2\cdot\hat{\bm{k}})\right)\sin^2\phi,\nonumber\\
(\hat{\bm{q}}_1\cdot\hat{\bm{y}})^2+(\hat{\bm{q}}_2\cdot\hat{\bm{y}})^2&=\left(2-(\hat{\bm{q}}_1\cdot\hat{\bm{k}})^2-(\hat{\bm{q}}_2\cdot\hat{\bm{k}})^2\right)\sin^2\phi.
\end{align}
Now we can rewrite $h_E$ as
\begin{align}
h_E(\hat{\bm{q}}_1,\hat{\bm{q}}_2)=&(\hat{\bm{q}}_1\cdot\hat{\bm{q}}_2)(\hat{\bm{q}}_1\cdot\hat{\bm{k}})(\hat{\bm{q}}_2\cdot\hat{\bm{k}})(1+\sin^2\phi)-(\hat{\bm{q}}_1\cdot\hat{\bm{q}}_2)^2\sin^2\phi\nonumber\\
 &-\frac{1}{3}(1+\sin^2\phi)\left[(\hat{\bm{q}}_1\cdot\hat{\bm{k}})^2+(\hat{\bm{q}}_2\cdot\hat{\bm{k}})^2\right]+\frac{2}{3}\sin^2\phi\nonumber\\
=&\mu\mu_1\mu_2(1+\sin^2\phi)-\mu^2\sin^2\phi-\frac{1}{3}(1+\sin^2\phi)(\mu_1^2+\mu_2^2)+\frac{2}{3}\sin^2\phi,
\end{align}
where we define $\mu\equiv\hat{\bm{q}}_1\cdot\hat{\bm{q}}_2,\mu_1\equiv\hat{\bm{q}}_2\cdot\hat{\bm{k}},\mu_2\equiv\hat{\bm{q}}_1\cdot\hat{\bm{k}}$ (following the convention where each angle is labeled by the subscript for the opposite side in the triangle).

Taking square of $h_E$ and then averaging over $\phi$, we obtain\footnote{Averaging over the azimuthal angle, we have $\langle\cos^2\phi\rangle=\frac{1}{2\pi}\int_0^{2\pi}d\phi\,\cos^2\phi=1/2$, $\langle\cos^4\phi\rangle=\frac{1}{2\pi}\int_0^{2\pi}d\phi\,\cos^4\phi=3/8$. More generally, $\langle\cos^{2n}\phi\rangle=\pi^{-\frac{1}{2}}\Gamma(n+\frac{1}{2})/\Gamma(n+1)$ for any non-negative integer $n$, known as the Wallis formula.
}
\begin{align}
h_E^2=&\frac{1}{6}-\frac{1}{2}\mu^2+\frac{3}{8}\mu^4-\frac{7}{18}(\mu_1^2+\mu_2^2)+\frac{7}{12}\mu^2(\mu_1^2+\mu_2^2)+\frac{19}{72}(\mu_1^4+\mu_2^4)\nonumber\\
&+\frac{7}{6}\mu\mu_1\mu_2-\frac{7}{4}\mu^3\mu_1\mu_2-\frac{19}{12}\mu(\mu_1^3\mu_2+\mu_1\mu_2^3)+\frac{19}{36}\mu_1^2\mu_2^2+\frac{19}{8}\mu^2\mu_1^2\mu_2^2\nonumber\\
=&\sum_{\substack{\ell_1,\ell_2,\ell\\ \ell_1\geq\ell_2}}A_{\ell_1\ell_2\ell}^{00(E)}~\mathcal{P}_{\ell}(\mu)\mathcal{P}_{\ell_1}(\mu_1)\mathcal{P}_{\ell_2}(\mu_2)~,
\end{align}
where we apply the symmetry between $\bm{q}_1$ and $\bm{q}_2$ and only keep terms with $\ell_1\geq\ell_2$. Similarly, we can write $h_B^2$ kernel in the same form with coefficients $A_{\ell_1\ell_2\ell}^{00(B)}$. The coefficient of each term is listed in Table \ref{tab:coefficients}. Now each term has been expressed in the required form of $q_1^\alpha q_2^\beta\mathcal{P}_{\ell}(\mu)\mathcal{P}_{\ell_1}(\mu_1)\mathcal{P}_{\ell_2}(\mu_2)$, with $\alpha=\beta=0$.

\begin{table}
\centering
\begin{tabular}{| c | c | c | c | c |}
 \hline
 \rule{0pt}{2.5ex} $\ell$ & $\ell_1$ & $\ell_2$ & $A_{\ell_1\ell_2\ell}^{00(E)}$ & $A_{\ell_1\ell_2\ell}^{00(B)}$\\
 \hline
 $0$& $0$ & $0$ & $\nicefrac{16}{81}$ & $-\nicefrac{41}{405}$\\
       & $2$ & $0$ & $\nicefrac{713}{1134}$ & $-\nicefrac{298}{567}$ \\
       & $2$ & $2$ & $\nicefrac{95}{162}$ & $-\nicefrac{40}{81}$ \\
       & $4$ & $0$ & $\nicefrac{38}{315}$ & $-\nicefrac{32}{315}$\\
 \hline
 $1$& $1$ & $1$ & $-\nicefrac{107}{60}$ & $\nicefrac{59}{45}$  \\
       & $3$ & $1$ & $-\nicefrac{19}{15}$ & $\nicefrac{16}{15}$  \\
 \hline
 $2$& $0$ & $0$ & $\nicefrac{239}{756}$ & $-\nicefrac{2}{9}$ \\
       & $2$ & $0$ & $\nicefrac{11}{9}$ & $-\nicefrac{20}{27}$ \\
       & $2$ & $2$ & $\nicefrac{19}{27}$ & $-\nicefrac{16}{27}$ \\
 \hline
 $3$& $1$ & $1$ & $-\nicefrac{7}{10}$ & $\nicefrac{2}{5}$ \\
 \hline
 $4$& $0$ & $0$ & $\nicefrac{3}{35}$ & --- \\
 \hline
\end{tabular}
\caption{The coefficient of each term in the Legendre polynomial expansion of $h_E^2$ and $h_B^2$ kernels (without the factor of 2 in front of the integral Eq.~\ref{eq:Pk-IA}). Due to symmetry, we need only keep terms with $\ell_1\geq \ell_2$ (multiplying the value by two where relevant).}
\label{tab:coefficients}
\end{table}

\begin{figure}
\centering
\includegraphics[scale=0.38]{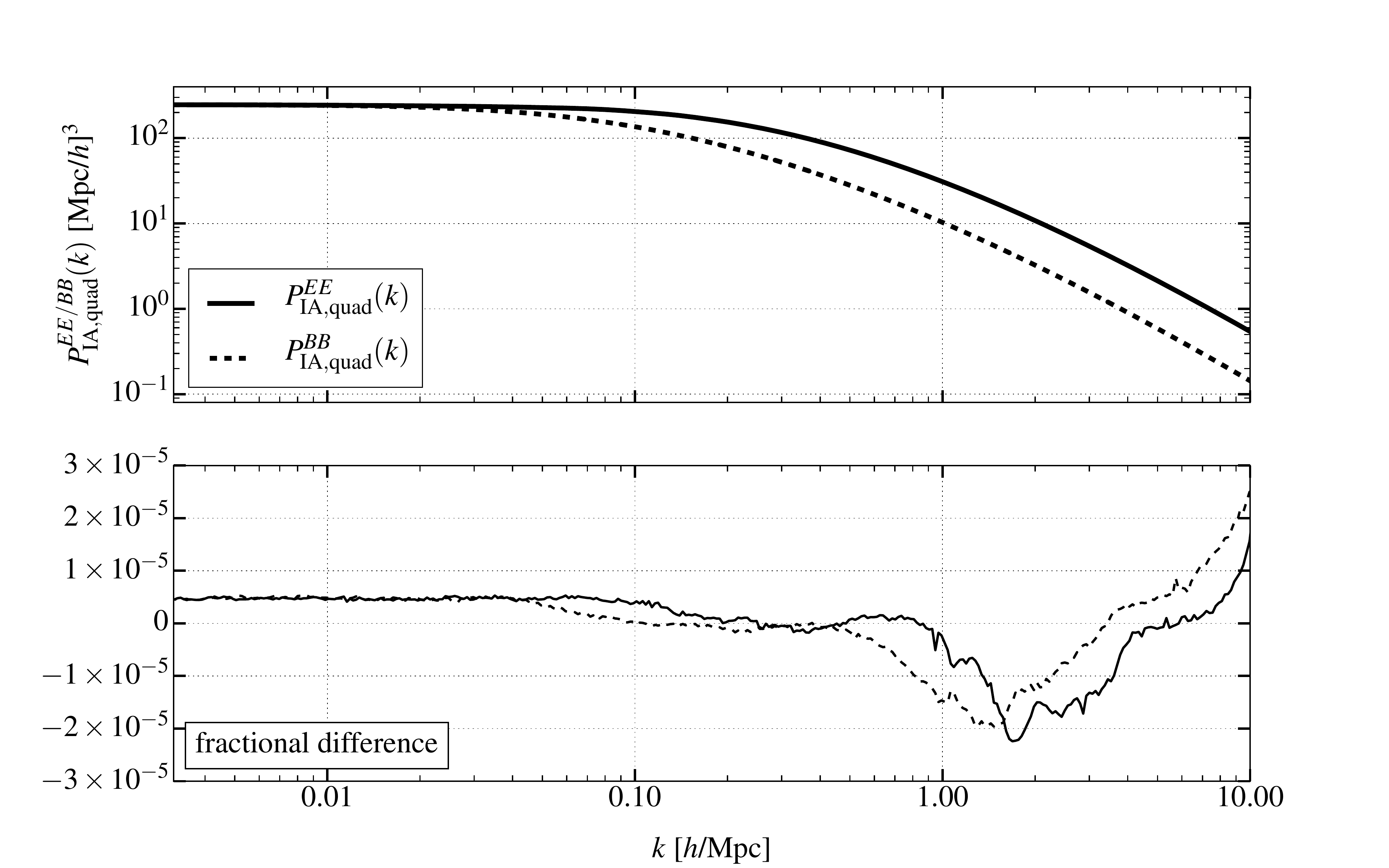}
\caption{The {\py FAST-PT} result for the intrinsic alignment integrals $P_{\rm IA,quad}^{(EE,BB)}(k)$ in Eq. (\ref{eq:Pk-IA}) (upper panel)  and the fractional difference compared to the conventional method (lower panel).}
\label{fig:IA}
\end{figure}

In Figure~\ref{fig:IA}, we show the {\py FAST-PT} result of $P_{\rm IA,quad}^{(EE,BB)}(k)$ (Eq. \ref{eq:Pk-IA}) and the fractional difference comparing to the results from conventional methods. The plot shows excellent agreement between two methods, with fractional accuracy better than $3\times 10^{-5}$ up to $k=10~h/$Mpc.

\subsection{Ostriker-Vishniac Effect}\label{subsec:OV}

\subsubsection{Theory}\label{subsub-OV-theory}

After CMB photons leave the surface of last scattering, they can experience further interactions, leading to secondary anisotropies. One of the most important is re-scattering off of free electrons after reionization in which photons can be shifted to higher or lower frequencies due to motions of the electrons. The thermal Sunyaev-Zel'dovich effect (tSZ) results from thermal motion of the electrons, usually in galaxy clusters as these are the hottest regions. Bulk hydrodynamic motions produce the kinetic Sunyaev-Zel'dovich (kSZ) effect (in clusters) or the Ostriker-Vishniac (OV) effect (in large-scale structure). In this section, we consider the second-order perturbation theory analysis of the Ostriker-Vishniac effect.

The fractional temperature perturbation in the direction $\hat{\bm{n}}$ on the sky is given by \cite{1986ApJ...306L..51O,vishniac1987reionization,Jaffe:1997ye}
\begin{equation}
\Theta(\hat{\bm{n}})=-\int_0^{\eta_0}dw\,g(w)\hat{\bm{n}}\cdot\bm{q}(\bm{w})~,
\end{equation}
where $\bm{q}(\bm{w})\equiv[1+\delta(\bm{w})]\bm{v}(\bm{w})$, $\bm{v}(\bm{w})$ is the bulk velocity at position $\bm{w}\equiv w\hat{\bm{n}}$ at a comoving distance $w$ (or a conformal time $\eta_0-w$), $g(w)$ is the visibility function specifying the probability distribution for scattering from reionized electrons, given by $g(w)=(d\tau/dw)e^{-\tau}$, and $\tau$ is the optical depth.

At 1-loop, the angular power spectrum of $\Theta$ produced by the OV effect, $C^{\Theta\Theta}_{\ell}$ (equivalent to $P_p(\kappa)$ in \cite{Jaffe:1997ye}), requires the calculation of the Vishniac power spectrum, which is a tensor convolution integral. In a flat Universe,
\begin{equation}
C^{\Theta\Theta}_{\ell}=\frac{1}{16\pi^2}\int_0^{\eta_0}\frac{(a(w)g(w))^2}{w^2}\left(\frac{\dot{D}D}{D_0}\right)^2 S(\ell/w)dw~,
\end{equation}
where $D$ and $D_0$ are the growth factors at $w$ and at present, respectively. Choosing the same coordinate system as in the IA calculation above, \ie $\hat{\bm{z}}=\hat{\bm{k}}$ and $\hat{\bm{x}}$ pointing to the observer, the integral is given by
\begin{equation}
S(k)=4\pi^2\int\dq{1}\left(\frac{q_{1x}}{q_1^2}+\frac{q_{2x}}{q_2^2}\right)^2\Pl(q_1)\Pl(q_2)~,
\label{eq:Sk-OV}
\end{equation}
which is consistent with Eq. (21) in \cite{Jaffe:1997ye}. Our interest here is in fast computation of $S(k)$.

\subsubsection{Conversion to {\py FAST-PT} Format}\label{subsub-OV-conver}

First noting that the integral $S(k)$ is symmetric under the exchange $\bm{q}_1\leftrightarrow\bm{q}_2$ and that $q_{2x}=-q_{1x}$, we can expand Eq. (\ref{eq:Sk-OV}) as
\begin{equation}
S(k)=4\pi^2\int\dq{1}\left(\frac{2q_{1x}^2}{q_1^4}-\frac{2q_{1x}^2}{q_1^2q_2^2}\right)\Pl(q_1)\Pl(q_2)~.
\end{equation}
In the spherical coordinate system, $q_{1x}^2=q_1^2\sin^2\theta\cos^2\phi$, which becomes $\frac{1}{2}q_1^2\sin^2\theta$ after averaging over $\phi$. The kernel is thus
\begin{equation}
\left(1-(\hat{\bm{k}}\cdot\hat{\bm{q}}_1)^2\right)\left(\frac{1}{q_1^2}-\frac{1}{q_2^2}\right)=\frac{2}{3}[\mathcal{P}_0(\mu_2)-\mathcal{P}_2(\mu_2)]\left(\frac{1}{q_1^2}-\frac{1}{q_2^2}\right)~,
\end{equation}
where $\mu_2\equiv\hat{\bm{k}}\cdot\hat{\bm{q}}_1~$. There are 4 terms in this case: $A_{000}^{-2,0}=2/3$, $A_{020}^{-2,0}=-2/3$, $A_{000}^{0,-2}=-2/3$, $A_{020}^{0,-2}=2/3$.\footnote{It is possible to write the integral $S(k)$ in other forms without breaking the $\bm{q}_1\leftrightarrow\bm{q}_2$ symmetry, \eg to write the kernel as
$\left(1-\mu_2^2\right)\left(\frac{1}{2q_1^2}-\frac{1}{q_2^2}+\frac{q_1^2}{2q_2^4}\right)$. However, the $q_2^{-4}$ terms suffer from divergence at small $q_2$ (see \S\ref{subsec:div}). The divergence is artificial because $1-\mu_2^2\rightarrow 0$ when $q_2\rightarrow 0$, which makes physical sense, but it can cause instability in the {\py FAST-PT} code.}

\begin{figure}
\centering
\includegraphics[scale=0.38]{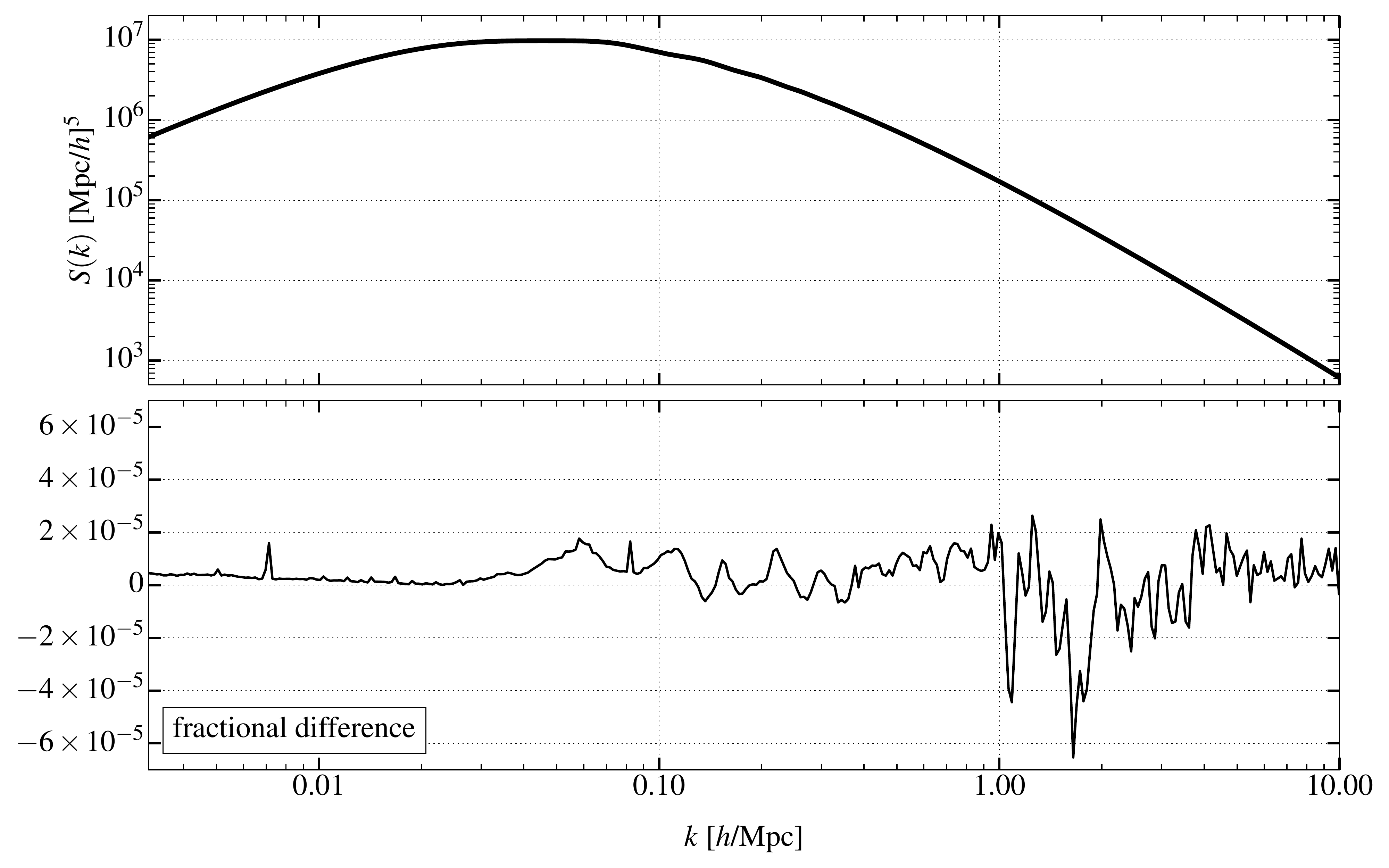}
\caption{The {\py FAST-PT} result for the Ostriker-Vishniac effect integral $S(k)$ in Eq. (\ref{eq:Sk-OV}) (upper panel)  and the fractional difference compared to the conventional method (lower panel).}
\label{fig:OV}
\end{figure}

In Figure~\ref{fig:OV}, we show the {\py FAST-PT} result of $S(k)$ integral (Eq. \ref{eq:Sk-OV}) and the fractional difference from a conventional method. The plot shows excellent agreement between two methods with accuracy better than $6\times 10^{-5}$ up to $k=10\ h/$Mpc.

\subsection{Kinetic polarization of the CMB}\label{subsec:kP}

\subsubsection{Theory}\label{subsub-kp-theory}

The kSZ effect can induce a secondary linear polarization in the CMB via the quadratic Doppler effect and Thomson scattering \cite{1980MNRAS.190..413S,Hu:1999vq}. Due to the motion of baryons, an isotropic CMB appears to have a quadrupole anisotropy component in the rest frame of the scattering baryons, as seen from the expansion
\begin{equation}
\Theta=\frac{\sqrt{1-v_b^2}}{1-\hat{\bm{n}}\cdot\bm{v}_b}-1\simeq \hat{\bm{n}}\cdot\bm{v}_b+\left(\hat{\bm{n}}\cdot\bm{v}_b\right)^2-\frac{1}{2}v_b^2~,
\end{equation}
where $\Theta$ is the fractional temperature fluctuation of CMB in the direction of $\hat{\bm{n}}$ as seen by the scattering electron. The relation between the quadrupole anisotropy at position $\bm{x}$ and the CMB temperature angular distribution seen by the scatter is given by
\begin{equation}
Q^{(m)}(\bm{x})=-\int d\Omega\frac{Y_{2m}^*(\hat{\bm{n}})}{\sqrt{4\pi}}\Theta(\bm{x},\hat{\bm{n}})~,
\label{eq:Qm}
\end{equation}
where $m=0,\pm 1,\pm 2$. In the Rayleigh-Jeans limit,\footnote{This limit is necessary to justify saying that {\em temperature} is scattered -- really it is the intensity, but at low frequencies the two are proportional. As noted in Ref.~\cite{1980MNRAS.190..413S}, the kinetic polarization has a specific non-blackbody spectral shape, which can be used to scale from the Rayleigh-Jeans limit to any frequency of interest.} the observed power spectra of $E$- and $B$-mode polarizations are related to the power spectra of $Q^{(0,\pm 2)}$ and $Q^{(\pm 1)}$, respectively, by
\begin{eqnarray}
C^{EE}_{\ell}&=&\frac{3\pi^2}{10\ell^3}\int dw\ g^2 D_A\left(\frac{3}{4}\Delta_Q^{2\ (0)}(k)+\frac{1}{8}\sum_{m=\pm 2}\Delta_Q^{2\ (m)}(k)\right)~~{\rm and} \nonumber \\
C^{BB}_{\ell}&=&\frac{3\pi^2}{10\ell^3}\int dw\ g^2 D_A\left(\frac{1}{2}\sum_{m=\pm 1}\Delta_Q^{2\ (m)}(k)\right)~,
\end{eqnarray}
where $\Delta_Q^{2\ (m)}(k) = k^3P^{(m)}(k)/(2\pi^2)$ is the variance of $Q^{(m)}$ per unit range in $\ln k$, the spherical harmonics in Eq.~(\ref{eq:Qm}) are evaluated with $\bm{k}$ on the $z$-axis, $g$ is the visibility function, and the comoving angular distance $D_A=w$ (the comoving distance) in a flat Universe. Since the quadrupole anisotropy arises from the quadratic Doppler effect, in Fourier space with $\hat{\bm{k}}=\hat{\bm{z}}$, we have
\begin{equation}
\label{eq:Qk}
Q^{(m)}(\bm{k})=-\int d\Omega\frac{Y_{2m}^*(\hat{\bm{n}})}{\sqrt{4\pi}}\int\dq{1} \hat{\bm{n}}\cdot\bm{v}_b(\bm{q}_1)\hat{\bm{n}}\cdot\bm{v}_b(\bm{q}_2)~,
\end{equation}
where $\bm{v}_b$ is the baryon bulk velocity. In linear theory
\begin{equation}
\dot{\delta}=-\frac{\bm{\nabla}\cdot\bm{v}_b}{a}=fH\delta~,
\end{equation}
where $f\equiv d\ln G/d\ln a$ for growth factor $G$ and scale factor $a$. Taking the Fourier transform and assuming no vorticity, we obtain
\begin{equation}
\label{eq:vb}
\bm{v}_b(\bm{k})=iafH\frac{\delta(\bm{k})}{k}\hat{\bm{k}}\equiv iT\frac{\delta(\bm{k})}{k}\hat{\bm{k}}~.
\end{equation}
Substituting Eq. (\ref{eq:vb}) into Eq. (\ref{eq:Qk}) and applying identities (\ref{eq:legen_add}, \ref{eq:3Y_integral}), we have
\begin{align}
Q^{(m)}(\bm{k})&=\frac{T^2}{\sqrt{4\pi}}\int\dq{1}\frac{\delta(\bm{q}_1)\delta(\bm{q}_2)}{q_1q_2} \int d\Omega~Y_{2m}^*(\hat{\bm{n}})\left(\hat{\bm{n}}\cdot\hat{\bm{q}}_1\right)\left(\hat{\bm{n}}\cdot\hat{\bm{q}}_2\right)\nonumber\\
&=\frac{T^2}{\sqrt{4\pi}}\int\dq{1}\frac{\delta(\bm{q}_1)\delta(\bm{q}_2)}{q_1q_2}\int d\Omega Y_{2m}^*(\hat{\bm{n}})\left(\frac{4\pi}{3}\right)^2\sum_{m_1m_2}Y_{1m_1}(\hat{\bm{q}}_1)Y_{1m_1}^*(\hat{\bm{n}})Y_{1m_2}(\hat{\bm{q}}_2)Y_{1m_2}^*(\hat{\bm{n}})\nonumber\\
&=\frac{T^2}{\sqrt{4\pi}}\int\dq{1}\frac{\delta(\bm{q}_1)\delta(\bm{q}_2)}{q_1q_2}\left(\frac{4\pi}{3}\right)^2\sum_{m_1m_2}Y_{1m_1}(\hat{\bm{q}}_1)Y_{1m_2}(\hat{\bm{q}}_2)\sqrt{\frac{45}{4\pi}}\left(\begin{array}{ccc}
2&1&1\\0&0&0
\end{array}\right)
\left(\begin{array}{ccc}
2&1&1\\m&m_1&m_2
\end{array}\right)\nonumber\\
&=\frac{T^2}{\sqrt{4\pi}}\left(\frac{4\pi}{3}\right)^2\sqrt{\frac{3}{2\pi}}\int \dq{1}\frac{\delta(\bm{q}_1)\delta(\bm{q}_2)}{q_1q_2}\sum_{m_1m_2}Y_{1m_1}(\hat{\bm{q}}_1)Y_{1m_2}(\hat{\bm{q}}_2)\left(\begin{array}{ccc}
2&1&1\\m&m_1&m_2
\end{array}\right)~.
\end{align}
Following the definition that $\langle Q^{(m)}(\bm{k})Q^{(m)}(\bm{k}')\rangle=(2\pi)^3P_{Q^{(m)}}(k)\delta_D^3(\bm{k}+\bm{k}')$, we have
\begin{equation}
P^{(m)}(k)\equiv \frac{27P_{Q^{(m)}}(k)}{4(4\pi)^2 T^4}=\int \dq{1}\frac{\Pl(q_1)\Pl(q_2)}{q_1^2q_2^2}\left\vert\sum_{m_1m_2}Y_{1m_1}(\hat{\bm{q}}_1)Y_{1m_2}(\hat{\bm{q}}_2)\left(\begin{array}{ccc}
2&1&1\\m&m_1&m_2
\end{array}\right)\right\vert^2~,
\label{eq:Pmk-kpol}
\end{equation}
which is a tensor convolution integral in the form of Eq.~(\ref{eq:legendre}).

\subsubsection{Conversion to {\py FAST-PT} Format}\label{subsub-kp-conver}
Since $\bm{k}\parallel\hat{\bm{z}}$, the kernels for each $m$ can be written in terms of $\mu,\mu_1,\mu_2$. Note that $m_1,m_2$ can only be 0 or $\pm 1$, so we can explicitly write down all the spherical harmonics and Wigner $3j$ symbols in the summation and transform to Legendre polynomial products as before:
\begin{align}
m=0:~&~~~\frac{q_1^{-2}q_2^{-2}}{80\pi^2}[2+6\mathcal{P}_2(\mu_1)+\mathcal{P}_2(\mu)+6\mathcal{P}_2(\mu_1)\mathcal{P}_2(\mu_2)-9\mathcal{P}_1(\mu_1)\mathcal{P}_1(\mu_2)\mathcal{P}_1(\mu)]~,\notag\\
m=\pm 1:~&~~~\frac{q_1^{-2}q_2^{-2}}{160\pi^2}[1-2\mathcal{P}_2(\mu_1)+9\mathcal{P}_1(\mu_1)\mathcal{P}_1(\mu_2)\mathcal{P}_1(\mu)-8\mathcal{P}_2(\mu_1)\mathcal{P}_2(\mu_2)]~,\notag\\
m=\pm 2:~&~~~\frac{q_1^{-2}q_2^{-2}}{80\pi^2}[1-2\mathcal{P}_2(\mu_1)+\mathcal{P}_2(\mu_1)\mathcal{P}_2(\mu_2)]~.
\end{align}
Note that the symmetry between $\mu_1$ and $\mu_2$ has been used to simplify the kernels. The coefficients $A_{\ell_1\ell_2\ell}^{\alpha\beta}$ are now trivially seen.

\begin{figure}
\centering
\includegraphics[scale=0.39]{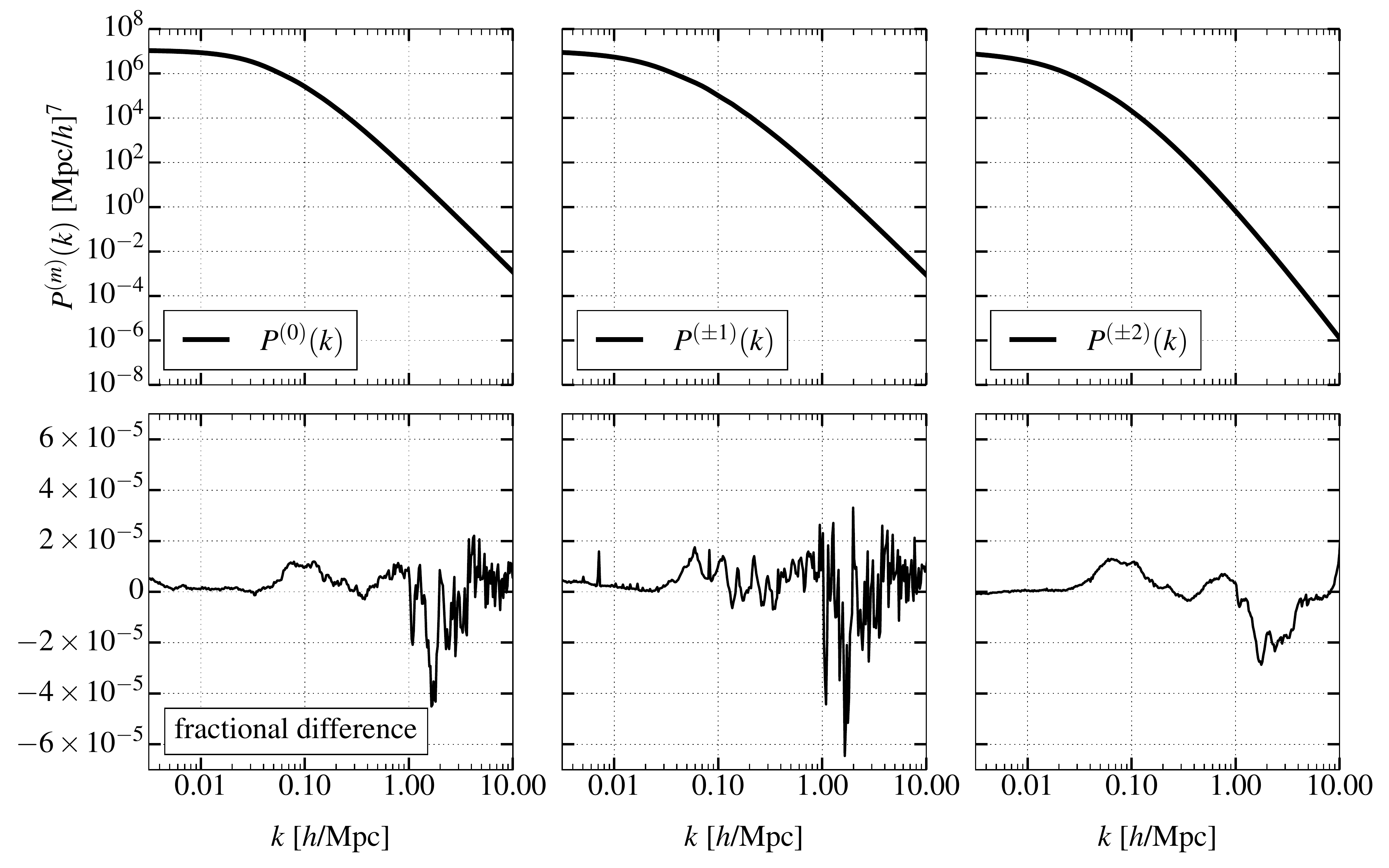}
\caption{The {\py FAST-PT} results for the kinectic CMB polarization integrals $P^{(m)}(k)$ in Eq. (\ref{eq:Pmk-kpol}) (upper panels)  and the fractional difference compared to the conventional method (lower panels).}
\label{fig:kpol}
\end{figure}

In Figure~\ref{fig:kpol}, we show the {\py FAST-PT} result of $P^{(m)}(k)$ integrals (Eq. \ref{eq:Pmk-kpol}) for $m=0,\pm 1,\pm 2$, respectively, and the fractional difference from a conventional method. The plots show excellent agreement between two methods with accuracy better than $6\times 10^{-5}$ in the $k$ range from $0.01$ to $10\ h/$Mpc.

\subsection{Redshift Space Distortions}\label{subsec:RSD}

\subsubsection{Theory}\label{subsub-RSD-theory}
Cosmological surveys map large-scale structure in three dimensions, using galaxies or other luminous tracers of the total matter distribution (e.g.\ \cite{Levi:2013gra,BOSS2013AJ....145...10D,EUCLID2011arXiv1110.3193L,WFIRST2013arXiv1305.5422S,Abbott:2016ktf}). To determine distance along the line-of-sight, surveys typically use redshift information and are thus actually making a map in ``redshift space.'' In order to compare theory to galaxy redshift survey data, models must be translated into redshift space.

Tracers tend to infall towards overdense regions and, due to the Doppler effect, will thus have observed redshifts that deviate from those predicted by pure cosmological expansion. These deviations cause ``redshift-space distortions'' (RSDs) in the observed tracer distribution. Although at highly nonlinear scales RSDs are no longer well-described by perturbation theory, \eg the ``Fingers of God'' (FoG) effect \cite{Jackson:2008yv}, we can still explore the mildly nonlinear regime via perturbation theory, avoiding time-consuming numerical simulations.

The ``textbook'' model for linear RSDs, the Kaiser effect \cite{1987MNRAS.227....1K}, relates the matter power spectrum in redshift space matter to that in real space matter with an angular-dependent bias factor related to the growth rate of structure. Subsequently, \cite{Scoccimarro:2004tg} improved the Kaiser model by distinguishing $P_{\delta\theta}$ and $P_{\theta\theta}$ from $P_{\delta\delta}$, where $\theta$ is the divergence of velocity field. In the linear regime of standard perturbation theory, these three power spectra are equal to each other.

The TNS model \cite{Taruya:2010mx} accounts for the nonlinear mode coupling between density and velocity fields, improving the modeling of the matter power spectrum in redshift space across a range of scales (including the BAO scale). Fixing $\bm{k}$ along the $\hat{\bm{z}}$ direction, and defining $\theta_n$ as the angle between $\hat{\bm{n}}$ (the line-of-sight direction) and $\bm{k}$, with $\mu_n\equiv\cos\theta_n$, the density power spectrum in the redshift space can be written:
\begin{equation}
P^{(S)}(k,\mu_n)=D_{\rm FoG}[k\mu_n f\sigma_v]\left\lbrace P_{\delta\delta}(k)+2f\mu_n^2 P_{\delta\theta}(k) +f^2\mu_n^4 P_{\theta\theta}(k)+A(k,\mu_n)+B(k,\mu_n)\right\rbrace~,
\label{eq:TNS}
\end{equation}
where $D_{\rm FoG}[k\mu_n f\sigma_v]$ encapsulates the contribution from the FoG effect. The $A,B$ terms are tensor convolution integrals given by
\begin{align}
\bar{A}(k,\mu_n)&\equiv \frac{A(k,\mu_n)}{k\mu_n f}=\int\dq{1}\frac{q_{1n}}{q_1^2}[B_\sigma(\bm{q}_1,\bm{q}_2,-\bm{k})-B_\sigma(\bm{q}_1,\bm{k},-\bm{k}-\bm{q}_1)]~,\\
\bar{B}(k,\mu_n)&\equiv \frac{B(k,\mu_n)}{(k\mu_n f)^2}=\int\dq{1}F(\bm{q}_1)F(\bm{q}_2)~,
\end{align}
where $\bm{k}=\bm{q}_1+\bm{q}_2$, and the subscript ``$n$'' denotes the projection onto $\hat{\bm{n}}$, \eg $q_{1n}\equiv\bm{q}_1\cdot\hat{\bm{n}}$, and
\begin{equation}
F(\bm{q})=\frac{q_n}{q^2}\left(P_{\delta\theta}(q)+f\frac{q_n^2}{q^2}P_{\theta\theta}(q)\right)~.
\end{equation}
The cross bispectra $B_\sigma$ is defined by
\begin{equation}
\left\langle\theta(\bm{k}_1)\left[\delta(\bm{k}_2)+f\frac{k_{2n}^2}{k_2^2}\theta(\bm{k}_2)\right]\left[\delta(\bm{k}_3)+f\frac{k_{3n}^2}{k_3^2}\theta(\bm{k}_3)\right]\right\rangle=(2\pi)^3\delta_D(\bm{k}_1+\bm{k}_2+\bm{k}_3)B_\sigma(\bm{k}_1,\bm{k}_2,\bm{k}_3)~.
\label{eq:Bsigma}
\end{equation}
The convolution integrals $\bar{A}(k,\mu_n)$ and $\bar{B}(k,\mu_n)$ are particularly time-consuming (\eg \cite{Bose:2016qun}) and are ideal applications for our algorithm.

\paragraph{$\bar{B}$ Term\\}
Substituting the $F(\bm{q})$ kernel into the $\bar{B}(k,\mu_n)$ integral, we obtain
\begin{align}
\bar{B}(k,\mu_n)=& \int \dq{1}\frac{q_{1n}q_{2n}}{q_1^2 q_2^2}\left[P_{\delta\theta}(q_1) P_{\delta\theta}(q_2) + f^2\frac{q_{1n}^2 q_{2n}^2}{q_1^2q_2^2}P_{\theta\theta}(q_1)P_{\theta\theta}(q_2)+ 2f\frac{q_{2n}^2}{q_2^2}P_{\delta\theta}(q_1) P_{\theta\theta}(q_2)\right]\nonumber\\
=& \int\dq{1}\frac{(\hat{\bm{q}}_1\cdot\hat{\bm{n}})(\hat{\bm{q}}_2\cdot\hat{\bm{n}})}{q_1q_2} \Pl(q_1)\Pl(q_2) + f^2\int\dq{1}\frac{(\hat{\bm{q}}_1\cdot\hat{\bm{n}})^3 (\hat{\bm{q}}_2\cdot\hat{\bm{n}})^3}{q_1q_2}\Pl(q_1) \Pl(q_2)\nonumber\\
& + 2f\int\dq{1}\frac{(\hat{\bm{q}}_1\cdot\hat{\bm{n}}) (\hat{\bm{q}}_2\cdot\hat{\bm{n}})^3}{q_1q_2}\Pl(q_1) \Pl(q_2)~.
\end{align}
As previously mentioned, $P_{\delta\theta},P_{\theta\theta},P_{\delta\delta}$ are all equal to $\Pl$ at the leading order. Since terms in the form of $(\hat{\bm{q}}_1\cdot\hat{\bm{n}})^{p_1}(\hat{\bm{q}}_2\cdot\hat{\bm{n}})^{p_2}$ with non-negative integers $p_1,p_2$ can always be decomposed as a polynomial in terms of $\hat{\bm{k}}\cdot\hat{\bm{n}}$ after longitude angle averaging (see Appendix \ref{app:proof} for a proof), it is natural to write $\bar{B}$ as
\begin{equation}
\bar{B}(k,\mu_n)=\sum_{i=0}B_i(k)\mu_n^i~,
\end{equation}
where each $B_i(k)$ is a tensor convolution integral that can be written in terms of products of Legendre polynomials.

\paragraph{$\bar{A}$ Term\\}
The cross bispectrum satisfies $B_\sigma(\bm{k}_1,\bm{k}_2,\bm{k}_3)=B_\sigma(-\bm{k}_1,-\bm{k}_2,-\bm{k}_3)=B_\sigma(\bm{k}_1,\bm{k}_3,\bm{k}_2)$, so we can write the $\bar{A}$ integral as
\begin{equation}
\bar{A}(k,\mu_n)=\int\dq{1}\frac{q_{1n}}{q_1^2}[B_\sigma(\bm{q}_1,\bm{q}_2,-\bm{k})-B_\sigma(-\bm{q}_1,\bm{k}+\bm{q}_1,-\bm{k})]~.
\end{equation}
Changing the dummy variable $\bm{q}_1$ to $-\bm{q}_1$ in the second term, we have
\begin{equation}
\bar{A}(k,\mu_n)=\int\dq{1}\frac{q_{1n}}{q_1^2}[B_\sigma(\bm{q}_1,\bm{q}_2,-\bm{k})+B_\sigma(\bm{q}_1,\bm{k}-\bm{q}_1,-\bm{k})]=2\int\dq{1}\frac{q_{1n}}{q_1^2}B_\sigma(\bm{q}_1,\bm{q}_2,-\bm{k})~.
\end{equation}
Expanding the left-hand side of Eq. (\ref{eq:Bsigma}) to the leading order, we have
\begin{align}
B_\sigma(\bm{q}_1,\bm{q}_2,-\bm{k})=& 2\left(1+\frac{q_{2n}^2}{q_2^2}f\right)\left(1+\frac{k_{n}^2}{k^2}f\right) G_2(\bm{q}_2,-\bm{k})\Pl(q_2)\Pl(k)\nonumber\\
& + 2\left(1+\frac{k_{n}^2}{k^2}f\right)\left(F_2(\bm{q}_1,-\bm{k})+\frac{q_{2n}^2}{q_2^2}fG_2(\bm{q}_1,-\bm{k})\right)\Pl(q_1)\Pl(k)\nonumber\\
& + 2\left(1+\frac{q_{2n}^2}{q_2^2}f\right)\left(F_2(\bm{q}_1,\bm{q}_2)+\frac{k_{n}^2}{k^2}fG_2(\bm{q}_1,\bm{q}_2)\right)\Pl(q_1)\Pl(q_2)~.
\end{align}
Similarly, we can expand the integral $\bar{A}$ as a polynomial in terms of $\mu_n$:
\begin{equation}
\bar{A}(k,\mu_n)=\sum_{i=0}A_i(k)\mu_n^i~.
\end{equation}
Each $A_i(k)$ can be separated into two parts:
\begin{equation}
A_i(k)=A^{\rm I}_i(k)+A^{\rm II}_i(k)~,
\end{equation}
where $A^{\rm I}_i(k)$ is a convolution integral with $\Pl(q_1)\Pl(q_2)$ in the integrand, while $A^{\rm II}_i(k)$ has an integrand with $\Pl(q_1)\Pl(k)$ or $\Pl(q_2)\Pl(k)$, which is similar to the $P_{13}$ integral and can be treated in a similar fashion.

\subsubsection{Conversion to {\py FAST-PT} Format}\label{subsub-RSD-conver}
The $B_i(k)$ and $A_i^{\rm I}(k)$ integrals are standard convolution integrals, which can be decomposed into the form of Eq.~(\ref{eq:legendre}). The associated coefficients $A_{\ell_1\ell_2\ell}^{\alpha\beta}$ are listed in Tables \ref{tab:B-term} and \ref{tab:A-term}.

\begin{table}
\centering
\begin{tabular}{| c |c | c | c | c | c | c | c |}
 \hline
\multirow{2}{*}{ }& \multirow{2}{*}{$\ell$} & \multirow{2}{*}{$\ell_1$} & \multirow{2}{*}{$\ell_2$} & \multicolumn{4}{ c| }{$A_{\ell_1\ell_1\ell}^{\alpha\beta}$} \\ \cline{5-8}
                & & & & $i=0$ &$i=2$&$i=4$&$i=6$ \\
 \hline
$B_i$&$0$& $1$ & $1$ & $-\nicefrac{1}{2}-\nicefrac{3f}{10}-\nicefrac{f^2}{20}$&$\nicefrac{3}{2}+\nicefrac{3f^2}{20}$&$\nicefrac{3f}{2}-\nicefrac{21f^2}{20}$&$\nicefrac{131f^2}{100}$\\
         &       & $3$ & $1$ & $\nicefrac{3f}{10}+\nicefrac{f^2}{10}$& $-3f-\nicefrac{6f^2}{5}$&$\nicefrac{7f}{2}-\nicefrac{3f^2}{10}$&$\nicefrac{47f^2}{25}$ \\
         &       & $3$ & $3$ & $-\nicefrac{f^2}{20}$ &$\nicefrac{21f^2}{20}$& $-\nicefrac{63f^2}{20}$&$\nicefrac{231f^2}{100}$\\
         & $1$& $0$ & $0$ & $\nicefrac{(1+f)}{2}+\nicefrac{5f^2}{36}$&$-\nicefrac{1}{2}+\nicefrac{f^2}{12}$&$-\nicefrac{f}{2}-\nicefrac{f^2}{12}$& $-\nicefrac{5f^2}{36}$\\
         &       & $2$ & $0$ & $-\nicefrac{f}{2}-\nicefrac{5f^2}{18}$ &$3f+\nicefrac{4f^2}{3}$ &$-\nicefrac{5f}{2}+\nicefrac{f^2}{6}$&$-\nicefrac{11f^2}{9}$\\
         &       & $2$ & $2$ & $\nicefrac{5f^2}{36}$& $-\nicefrac{17f^2}{12}$&$\nicefrac{53f^2}{12}$&$-\nicefrac{113f^2}{36}$\\
 \hline
\end{tabular}
\caption{The coefficient of each term in the Legendre polynomial expansion of kernels of $B_i(k)$. $\alpha=\beta=-1$ for all the terms. Due to symmetry, we need only keep terms with $\ell_1\geq \ell_2$ (multiplying the value by two where relevant). Empty entries are equal to the previous row.}
\label{tab:B-term}
\end{table}

\begin{table}
\centering
\begin{tabular}{| c |c | c | c | c | c | c | c | c | c |}
 \hline
\multirow{2}{*}{ }& \multirow{2}{*}{$\alpha$}&\multirow{2}{*}{$\beta$}&\multirow{2}{*}{$\ell$} & \multirow{2}{*}{$\ell_1$} & \multirow{2}{*}{$\ell_2$} & \multicolumn{3}{ c| }{$A_{\ell_1\ell_1\ell}^{\alpha\beta}$} \\ \cline{7-9}
                & & & & & & $i=1$ &$i=3$&$i=5$ \\
 \hline
$A^{\rm I}_i$&$-1$&$0$&$0$& $0$ & $1$ & $\nicefrac{68}{21}+\nicefrac{2f}{3}$&$\nicefrac{26f}{9}+\nicefrac{2f^2}{3}$& $\nicefrac{10f^2}{63}$ \\
           &       &      &       & $2$ & $1$ & $-\nicefrac{68f}{21}$ &$\nicefrac{340f}{63}-\nicefrac{52f^2}{21}$&$\nicefrac{260f^2}{63}$\\
           &       &      &$1$ & $1$ & $0$ & $2+\nicefrac{124f}{35}$ &$-\nicefrac{92f}{105}+\nicefrac{108f^2}{35}$&$-\nicefrac{254f^2}{105}$\\
           &       &      &       &        & $2$ & $-2f$ & $\nicefrac{10f}{3}-2f^2$&$\nicefrac{10f^2}{3}$\\
           &       &      &$2$ & $0$ & $1$ & $\nicefrac{16}{21}+\nicefrac{4f}{3}$&$\nicefrac{4f}{9}+\nicefrac{4f^2}{3}$& $-\nicefrac{52f^2}{63}$\\     
           &       &      &       & $2$ & $1$ & $-\nicefrac{16f}{21}$&$\nicefrac{80f}{63}-\nicefrac{32f^2}{21}$& $\nicefrac{160f^2}{63}$\\
           &       &      &$3$ & $1$ & $0$ & $\nicefrac{16f}{35}$&$-\nicefrac{16f}{35}+\nicefrac{32f^2}{35}$&$-\nicefrac{32f^2}{35}$\\   
           &$-2$&$1$&$0$ & $1$ & $0$ & $\nicefrac{2f}{3}$&$-\nicefrac{2f}{3}+\nicefrac{2f^2}{3}$&$-\nicefrac{2f^2}{3}$\\
           &       &      &$1$ & $0$ & $1$ & $2$&$\nicefrac{8f}{3}$&$\nicefrac{2f^2}{3}$ \\   
           &       &      &       & $2$ & $1$ & $-2f$&$\nicefrac{10f}{3}-2f^2$& $\nicefrac{10f^2}{3}$\\   
           &       &      &$2$ & $1$ & $0$ & $\nicefrac{4f}{3}$&$-\nicefrac{4f}{3}+\nicefrac{4f^2}{3}$& $-\nicefrac{4f^2}{3}$\\   
 \hline
\end{tabular}
\caption{The coefficient of each term in the Legendre polynomial expansion of kernels of $A^{\rm I}_i(k)$. The empty entries mean that they equal to the previous row.}
\label{tab:A-term}
\end{table}

The $A_i^{\rm II}(k)$ integrals are first decomposed into the form of
\begin{equation}
P_{\ell_1\ell_2\ell}^{\alpha\beta\gamma}(k)=\int\dq{1} q_1^\alpha q_2^\beta k^\gamma \mathcal{P}_{\ell_1}(\hat{\bm{q}}_2\cdot\hat{\bm{k}})\mathcal{P}_{\ell_2}(\hat{\bm{q}}_1\cdot\hat{\bm{k}})\mathcal{P}_{\ell}(\hat{\bm{q}}_1\cdot\hat{\bm{q}}_2)\Pl(q_1)\Pl(k)~,
\end{equation}
with coefficients $A_{\ell_1\ell_2\ell}^{\alpha\beta\gamma}$ given by Table \ref{tab:Ap-term}, so that for each $A_i^{\rm II}$ integral,
\begin{equation}
A_i^{\rm II}(k)=\sum A_{\ell_1\ell_2\ell}^{\alpha\beta\gamma}P_{\ell_1\ell_2\ell}^{\alpha\beta\gamma}(k)~.
\end{equation}
Note that for $\Pl(q_2)\Pl(k)$ terms one can always exchange the indices (1$\leftrightarrow$2) of $q$ and $\ell$ in the integrand to recover the form above. For the special case that $\beta=\ell_1=\ell=0$ and $\ell_2\neq 0$, the integral vanishes. These $P_{13}$-like integrals can be further reduced to one-dimensional integrals and quickly calculated using discrete convolutions as done for $P_{13}$ in \cite{McEwen:2016fjn}.
\begin{align}
P_{\ell_1\ell_2\ell}^{\alpha\beta\gamma}(k)=&k^\gamma\Pl(k) \int\dq{1} q_1^\alpha q_2^\beta  \mathcal{P}_{\ell_1}\left(\frac{k-q_1\mu_2}{q_2}\right)\mathcal{P}_{\ell_2}(\mu_2)\mathcal{P}_{\ell}\left(\frac{k\mu_2-q_1}{q_2}\right)\Pl(q_1)\nonumber\\
=&\frac{k^\gamma\Pl(k)}{(2\pi)^2} \int_0^\infty dq_1\,q_1^{2+\alpha}\Pl(q_1) \int_{-1}^1 d\mu_2\,q_2^\beta  \mathcal{P}_{\ell_1}\left(\frac{k-q_1\mu_2}{q_2}\right)\mathcal{P}_{\ell_2}(\mu_2)\mathcal{P}_{\ell}\left(\frac{k\mu_2-q_1}{q_2}\right)~,
\end{align}
where $q_2=\sqrt{k^2+q_1^2-2kq_1\mu_2}$. The angular ($\mu_2$) integration can be performed analytically.\footnote{There are several ways to do this; a brute-force approach is to write $\mu_2$ in terms of $q_2$ (at fixed $k$ and $q_1$), which turns the integral into a linear combination of power laws in $q_2$.} Summing the components, we find:
\begin{equation}
A^{\rm II}_i(k)=\frac{k^2\Pl(k)}{672\pi^2}\int_0^\infty dr\Pl(kr)Z_i(r)~,~~i=1,3,5,
\label{eq:int-AII}
\end{equation}
where
\begin{align}
Z_1(r)=&\frac{18f}{r^2}-152-66f+(192-66f)r^2-(72-18f)r^4\nonumber\\
 &+\left[\frac{9f}{r^3}+\frac{36(1-f)}{r}-54(2-f)r+36(3-f)r^3-9(4-f)r^5\right]\ln\left\vert\frac{1-r}{1+r}\right\vert~,\\
Z_3(r)=&  \frac{18 f(1+ f)}{r^2} -370 f - 66 f^2+ (318 f - 
 66 f^2) r^2 -( 126 f - 18 f^2) r^4 \nonumber\\
 &+ \left[\frac{9 f(1+ f)}{r^3} + \frac{36 f(1 - f)}{r} -54f(3 - f) r + 36f(5 - f) r^3 -9f(7 - f) r^5\right]\ln\left\vert\frac{1-r}{1+r}\right\vert\nonumber~,\\
Z_5(r)=& \frac{18 f^2}{r^2} -218 f^2 + 126 f^2 r^2 - 54 f^2 r^4 + 
\left[\frac{9 f^2}{r^3} - 54 f^2 r + 72 f^2 r^3 - 27 f^2 r^5\right]\ln\left\vert\frac{1-r}{1+r}\right\vert~.\nonumber
\end{align}
The integral (\ref{eq:int-AII}) is a convolution. Upon making the substitution $r=e^{-s}$, Eq.~(\ref{eq:int-AII}) becomes 
\begin{align}
A^{\rm II}_i(k) & = \frac{k^2\Pl(k)}{672 \pi^2} \int_{- \infty}^\infty  ds \; e^{-s}  \Pl(e^{\log k - s})  Z_i(e^{-s})\nonumber \\
& = \frac{k^2\Pl(k)}{672 \pi^2} \int_{- \infty}^{\infty} ds \; G_i(s) F(\log k - s)  ~,
\end{align} 
where $G_i(s)\equiv e^{-s}Z_i(e^{-s})$ and $F(s)\equiv \Pl(e^{s})$. We can convert to a discrete convolution with the substitutions $ds \to \Delta$, $\log k_n=\log k_0+n\Delta$, and $s_m=\log k_0+m\Delta$ (where $k_0$ is the smallest value in the $k$ array):
\begin{align}
\begin{split}
\int_{- \infty}^{\infty} ds \; G_i(s) F(\log k - s)
& \rightarrow  \Delta \displaystyle \sum_{m=0}^{N-1}  G_i^D(m)  F^D(n-m) ~, 
\end{split}
\end{align}
where in the final line we define the discrete functions $G_i^D(m)\equiv G_i(s_m)$ and $F^D(m)\equiv F(m\Delta)$. We then have
\begin{align}
\begin{split}
A_i^{\rm II}(k_n) & = \frac{k_n^2\Pl(k_n)\Delta}{672 \pi^2 }   [G_i^D \otimes F^D][n]~,~~i=1,3,5.
\end{split}
\end{align}
Thus $A_i^{\rm II}(k)$, which at first appears to involve order $N^2$ steps (an integral over $N$ samples at each of $N$ output values $k_n$), can in fact be computed for all output $k_n$ in $\mathcal{O}(N\log N)$ steps\footnote{In principle, $N$ is the size of the input $k$ array. However, to suppress the possible ringing and alising effects, we need to apply appropiate window functions, zero-padding or extend the input power spectrum into a larger range. The true value of $N$ is usually a few times larger than the original value, depending on the user's inputs and options.}.

Note that some integrals $P_{\ell_1\ell_2\ell}^{\alpha\beta\gamma}(k)$ suffer from a divergence due to contributions from small-scales. When summing to get $A_i^{\rm II}$, the divergent parts cancel each other precisely. Taking $q_1$ to be large, so that $\bm{q}_2\rightarrow -\bm{q}_1$ and $\Pl(q_1)\propto q_1^{-3}$, we have
\begin{equation}
P_{\ell_1\ell_2\ell}^{\alpha\beta\gamma}(k) \rightarrow \frac{(-1)^{\ell+\ell_1}\delta_{\ell_1\ell_2} k^\gamma \Pl(k)}{(2\ell_1+1)2\pi^2}\int dq_1\,q_1^{2+\alpha+\beta}\Pl(q_1)\propto \int dq_1\,q_1^{\alpha+\beta-1},
\end{equation}
so that the divergence appears when $\ell_1=\ell_2$ and $\alpha+\beta\geq 0$. In Table \ref{tab:Ap-term} there are 5 terms that suffer from this divergence problem. However these divergences cancel in $A_i^{\rm II}$; in our case, the cancellation occurs when doing the sum over $(\alpha,\beta,\gamma,\ell_1,\ell_2,\ell)$ to derive $Z_i(r)$.

\begin{table}
\centering
\begin{tabular}{| c |c |c | c | c | c | c | c | c | c | c |}
 \hline
\multirow{2}{*}{ }&\multirow{2}{*}{$\gamma$}& \multirow{2}{*}{$\alpha$}&\multirow{2}{*}{$\beta$}&\multirow{2}{*}{$\ell$} & \multirow{2}{*}{$\ell_1$} & \multirow{2}{*}{$\ell_2$} & \multicolumn{3}{ c| }{$A_{\ell_1\ell_1\ell}^{\alpha\beta\gamma}$} \\ \cline{8-10}
             &   & & & & & & $i=1$ &$i=3$&$i=5$ \\
 \hline
$A^{\rm II}_i$&$0$&$-1$&$0$&$0$ & $2$ & $1$ & $-\nicefrac{108f}{35}$ &$\nicefrac{36f}{7}-\nicefrac{108f^2}{35}$&$\nicefrac{36f^2}{7}$\\
       &    &       &      &       &        & $3$ & $-\nicefrac{32f}{35}$& $\nicefrac{32f}{21}-\nicefrac{32f^2}{35}$& $\nicefrac{32f^2}{21}$\\
       &    &       &      & $1$   & $1$ & $0$ & $\nicefrac{52f}{21}$& $-\nicefrac{52f}{21}+\nicefrac{52f^2}{21}$& $-\nicefrac{52f^2}{21}$\\
       &    &       &      &       &        & $2$ & $\nicefrac{32f}{21}$& $-\nicefrac{32f}{21}+\nicefrac{32f^2}{21}$& $-\nicefrac{32f^2}{21}$\\
       &    &$0$&$-1$&$0$ & $1$ & $0$ & $\nicefrac{52}{21}-\nicefrac{32f}{105}$ &$\nicefrac{80f}{21}-\nicefrac{32f^2}{105}$&$\nicefrac{4f^2}{3}$\\
       &    &       &      &       &        & $2$ & $\nicefrac{32}{21}-\nicefrac{428f}{147}$&$\nicefrac{1012f}{147}-\nicefrac{428f^2}{147}$& $\nicefrac{788f^2}{147}$\\
       &    &       &      &       &        & $4$ & $-\nicefrac{192f}{245}$&$\nicefrac{64f}{49}-\nicefrac{192f^2}{245}$&$\nicefrac{64f^2}{49}$\\
        &   &       &      &$1$ & $0$ & $1$ & $\nicefrac{108f}{35}$&$-\nicefrac{108f}{35}+\nicefrac{108f^2}{35}$& $-\nicefrac{108f^2}{35}$\\ 
        &   &       &      &       &        & $3$ & $\nicefrac{32f}{35}$&$-\nicefrac{32f}{35}+\nicefrac{32f^2}{35}$&$-\nicefrac{32f^2}{35}$\\
       &$-1$&$0$&$0$&$0$ & $0$ & $0$ & $-\nicefrac{2}{3}$ &$-\nicefrac{8f}{9}$&$-\nicefrac{2f^2}{9}$\\
      &     &       &      &       & $2$ & $0$ & $\nicefrac{2f}{3}$ &$-\nicefrac{10f}{9}+\nicefrac{2f^2}{3}$&$-\nicefrac{10f^2}{9}$\\   
      &     &       &      &       &        & $2$ & $\nicefrac{4f}{3}$&$-\nicefrac{20f}{9}+\nicefrac{4f^2}{3}$&$-\nicefrac{20f^2}{9}$ \\
       &    &       &      &  $1$  & $1$ & $1$ & $-2f$ &$2f-2f^2$&$2f^2$\\
       &    &$1$&$-1$&$0$ & $1$ & $1$ & $-2+\nicefrac{4f}{5}$&$-4f+\nicefrac{4f^2}{5}$&$-2f^2$ \\
       &    &       &      &       &        & $3$ & $\nicefrac{6f}{5}$ &$-2f+\nicefrac{6f^2}{5}$&$-2f^2$\\
       &    &       &      &$1$ & $0$ & $0$ & $-\nicefrac{2f}{3}$ &$\nicefrac{2f}{3}-\nicefrac{2f^2}{3}$& $\nicefrac{2f^2}{3}$\\ 
       &    &       &      &       &        & $2$ & $-\nicefrac{4f}{3}$&$\nicefrac{4f}{3}-\nicefrac{4f^2}{3}$&$\nicefrac{4f^2}{3}$\\
      &$1$  &$-2$&$0$&$0$ & $0$ & $0$ & $-\nicefrac{2}{3}$ & $-\nicefrac{8f}{9}$&$-\nicefrac{2f^2}{9}$\\
      &     &       &      &       & $2$ & $0$ & $\nicefrac{2f}{3}$ &$-\nicefrac{10f}{9}+\nicefrac{2f^2}{3}$ &$-\nicefrac{10f^2}{9}$\\   
      &     &       &      &       &        & $2$ & $\nicefrac{4f}{3}$ &$-\nicefrac{20f}{9}+\nicefrac{4f^2}{3}$&$-\nicefrac{20f^2}{9}$\\
      &     &       &      &$1$ & $1$ & $1$ & $-2f$ &$2f-2f^2$&$2f^2$ \\
       &    &$-1$&$-1$&$0$ & $1$ & $1$ & $-2+\nicefrac{4f}{5}$&$-4f+\nicefrac{4f^2}{5}$& $-2f^2$\\
       &    &       &      &       &        & $3$ & $\nicefrac{6f}{5}$&$-2f+\nicefrac{6f^2}{5}$&$-2f^2$ \\
       &    &       &      &$1$ & $0$ & $0$ & $-\nicefrac{2f}{3}$ &$\nicefrac{2f}{3}-\nicefrac{2f^2}{3}$&$\nicefrac{2f^2}{3}$\\
       &    &       &      &       &        & $2$ & $-\nicefrac{4f}{3}$& $\nicefrac{4f}{3}-\nicefrac{4f^2}{3}$& $\nicefrac{4f^2}{3}$\\
 \hline
\end{tabular}
\caption{The coefficient of each term in the Legendre polynomial expansion of kernels of $A^{\rm II}_i(k)$.}
\label{tab:Ap-term}
\end{table}

\begin{figure}
\centering
\includegraphics[scale=0.38]{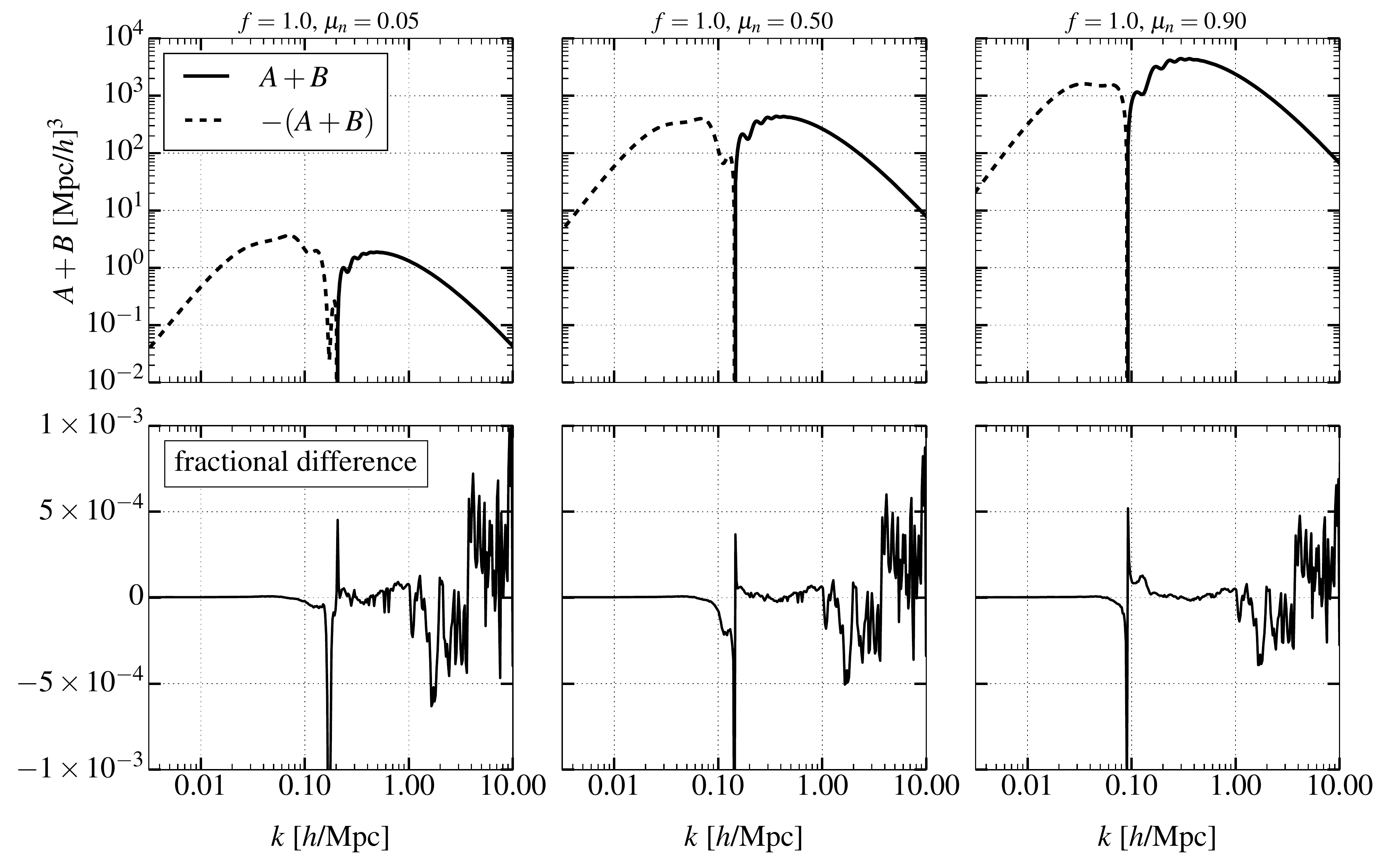}
\caption{The {\py FAST-PT} result for the redshift space distortion nonlinear corrections $A(k,\mu_n)+B(k,\mu_n)$ in the TNS model, Eq. (\ref{eq:TNS}) (upper panels)  and the fractional difference compared to the conventional method result (lower panels).}
\label{fig:RSDABsum}
\end{figure}

In Figures~\ref{fig:RSDABsum}, we show the {\py FAST-PT} results of $A+B$ terms in the TNS model (Eq. \ref{eq:TNS}) for $f=1$ and $\mu_n=0.05,0.5,0.9$, respectively, as well as the fractional difference compared to our conventional method. The plots show excellent agreement between two methods with accuracy at the $10^{-4}$ level for most of the $k$ range from $0.01$ to $10\ h/$Mpc. Note that the individual $A$ and $B$ terms agree to significantly higher precision ($\sim 10^{-5}$). Cancellations among terms in the total $A+B$ amplify the fractional difference, especially at high $k$ and near the zero-crossing.

\section{Summary}\label{sec:summary}
In this paper we have extended the {\py FAST-PT} algorithm to treat 1-loop convolution integrals with tensor kernels (explicitly dependent on the direction of the observed mode). The generalized algorithm has many applications -- we have presented quadratic intrinsic alignments, the Ostriker-Vishniac effect, kinetic CMB polarizations, and a sophisticated model for redshift space distortions. Our algorithm and code achieve high precision for all of these applications. We have tested the output of the code to high wavenumber ($k=10~h/$Mpc), although we reiterate that the smaller scales considered are beyond the range of validity of the underlying perturbative models. The reduction in evaluation time is similar as for the scalar {\py FAST-PT}. For instance, execution time is $\sim 0.1$ seconds for 600 $k$ values in all our examples. In the results shown here, the input power spectrum was sampled at 100 points per $\log_{10}$ interval. We find that much of the noise (in comparisons with the conventional method) is driven by the exact process by which the CAMB power spectrum is interpolated before it is used in {\py FAST-PT}.

There are underlying physical concepts and symmetries that make the efficiency of this algorithm possible. For example, the locality of the gravitational interactions allows us to separate different modes in configuration space. Since the structure evolution under gravity only depends on the local density and velocity divergence fields, in Fourier space the 1-loop power spectra of the matter density as well as its tracers (assuming local biasing theories) must be in form of Eq.~(\ref{eq:tensor_int}), where the kernels can always be written in terms of dot products of different mode vectors. Without this locality, it may not be possible to write the desired power spectrum as a sum of terms that can be calculated with this algorithm. The scale invariance of the problem also indicates that we should decompose the input power spectrum into a set of power-law spectra and make full use of the FFT algorithm. There are also rotational symmetries that allow us to reduce the 3-dimensional integrals to 1-dimension.

This algorithm, and implementations of the examples presented here, are publicly available as a Python code package at \url{https://github.com/JoeMcEwen/FAST-PT}.

\acknowledgments
XF is supported by the Simons Foundation, JB is supported by a CCAPP Fellowship, JM is supported by NSF grant AST1516997, and CH by the Simons Foundation, the US Department of Energy, the Packard Foundation, and NASA.

\bibliographystyle{JHEP.bst}
\bibliography{references}

\appendix
\section{Mathematical Identities}
In this work we have used a number of common mathematical identities. These identities are easily found in any standard mathematical physics text or handbook, (e.g. \cite{abramowitz1964handbook, NIST:DLMF, Olver:2010:NHMF}). However, to make our paper self-contained we list those relevant to our paper.
\subsection{Spherical Harmonics and Legendre Polynomials}
\begin{itemize}[leftmargin=*]
\item The addition theorem
\begin{align}
\label{eq:legen_add}
P_{\ell}(\hat{\bm{q}}_1 \cdot \hat{\bm{q}}_2) =\frac{4 \pi}{2\ell + 1} \displaystyle \sum_{m=-\ell}^\ell Y_{\ell m}(\hat{\bm{q}}_1) Y^*_{\ell m}(\hat{\bm{q}}_2);
\end{align}
\item The special case thereof,
\begin{align}
\label{eq:sph_add}
\displaystyle \sum_{m=-\ell}^\ell Y_{\ell m}(\hat{\bm{q}})Y^*_{\ell m}(\hat{\bm{q}}) = \frac{2\ell + 1}{4 \pi} ~;
\end{align}
\item The orthonormality relation
\begin{align} 
\label{eq:sph_ortho}
\int_{S^2} d^2 \hat{\bm{q}}\, Y_{\ell m}(\hat{\bm{q}}) Y_{\ell'm'}^*(\hat{\bm{q}}) = \delta_{\ell \ell'}\delta_{mm'} ~;
\end{align}
\item The symmetry
\begin{equation}
\label{eq:sph_sym}
Y_{\ell m}(\hat{\bm{q}})=(-1)^m~Y_{\ell,-m}^*(\hat{\bm{q}});
\end{equation}
\item The expansion/decomposition of a plane wave:
\begin{align}
\label{eq:planewave}
\int_{S^2} d^2 \hat{\bm{q}}\, Y^*_{\ell m}(\hat{\bm{q}}) e^{i \bm{q} \cdot \bm{r}} = 4 \pi  i^\ell j_{\ell}(qr) Y^*_{\ell m}(\hat{\bm{r}})
~~~\leftrightarrow~~~
 e^{i \bm{q} \cdot \bm{r}} = 4 \pi \displaystyle \sum_{\ell} \displaystyle i^\ell j_{\ell}(qr)  \sum_{m=-\ell}^\ell Y^*_{\ell m}(\hat{\bm{q}})Y_{\ell m}(\hat{\bm{r}})~.
\end{align}
\end{itemize}

\subsection{Wigner $3j$ and $6j$ Symbols}
The definitions of Wigner $3j$ and $6j$ symbols, denoted by ( ) and \{ \}, respectively, are long and can be easily found online or in handbooks. Here we only list some properties and identities needed in our derivations.
\begin{itemize}[leftmargin=*]
\item Assuming $j_1,j_2,j_3$ satisfy the triangle conditions, we have the special case
\begin{equation}
\left(\begin{array}{ccc}
j_1 & j_2 & j_3\\
0 & 0 & 0
\end{array}\right)
=\left\lbrace\begin{array}{ll}
0, & J~{\rm odd},\\
(-1)^{J/2}\left(\frac{(J-2j_1)!~(J-2j_2)!~(J-2j_3)!}{(J+1)!}\right)^{1/2}\frac{\left(\frac{1}{2}J\right)!}{\left(\frac{1}{2}J-j_1\right)!~\left(\frac{1}{2}J-j_2\right)!~\left(\frac{1}{2}J-j_3\right)!~}, & J~{\rm even},
\end{array}\right.
\end{equation}
where $J\equiv j_1+j_2+j_3~$;\\
\item The permutation and reflection symmetry
\begin{align}
\label{eq:3j_sym_permu}
\left(\begin{array}{ccc}
j_1 & j_2 & j_3\\
m_1 & m_2 & m_3
\end{array}\right)
&=
\left(\begin{array}{ccc}
j_2 & j_3 & j_1\\
m_2 & m_3 & m_1
\end{array}\right)
=
\left(\begin{array}{ccc}
j_3 & j_1 & j_2\\
m_3 & m_1& m_2
\end{array}\right)~,\nonumber\\
\left(\begin{array}{ccc}
j_1 & j_2 & j_3\\
m_1 & m_2 & m_3
\end{array}\right)
&=
(-1)^{j_1+j_2+j_3}\left(\begin{array}{ccc}
j_2 & j_1 & j_3\\
m_2 & m_1 & m_3
\end{array}\right)~;\\
\left(\begin{array}{ccc}
j_1 & j_2 & j_3\\
m_1 & m_2 & m_3
\end{array}\right)
&=
(-1)^{j_1+j_2+j_3}\left(\begin{array}{ccc}
j_1 & j_2 & j_3\\
-m_1 & -m_2 & -m_3
\end{array}\right)~;
\label{eq:3j_sym_ref}
\end{align}
\item The orthogonality relation
\begin{equation}
\label{eq:3j_ortho}
\sum_{m_{1}m_{2}}(2j_{3}+1)\begin{pmatrix}j_{1}&j_{2}&j_{3}\\
m_{1}&m_{2}&m_{3}\end{pmatrix}\begin{pmatrix}j_{1}&j_{2}&j^{\prime}_{3}\\
m_{1}&m_{2}&m^{\prime}_{3}\end{pmatrix}=\delta_{j_{3},j^{\prime}_{3}}\delta_{m%
_{3},m^{\prime}_{3}},
\end{equation}
\item Relation to spherical harmonics
\begin{equation}
\label{eq:Y_to_3j}
Y_{\ell_1m_1}(\hat{\bm{q}})Y_{\ell_2m_2}(\hat{\bm{q}})=\sum_{\ell,m}\sqrt{\frac{(2\ell_1+1)(2\ell_2+1)(2\ell+1)}{4\pi}}
\left(\begin{array}{ccc}
\ell_1 & \ell_2 & \ell\\
m_1 & m_2 & m
\end{array}\right)
Y_{\ell m}^*(\hat{\bm{q}})
\left(\begin{array}{ccc}
\ell_1 & \ell_2 & \ell\\
0 & 0 & 0
\end{array}\right);
\end{equation}
\begin{equation}
\label{eq:3Y_integral}
\int d^2\hat{\bm{q}}\mathop{Y_{{\ell_{1}}{m_{1}}}\/}\nolimits\!\left(%
\hat{\bm{q}}\right)\mathop{Y_{{\ell_{2}}{m_{2}}}\/}\nolimits\!\left(\hat{\bm{q}}%
\right)\mathop{Y_{{\ell_{3}}{m_{3}}}\/}\nolimits\!\left(\hat{\bm{q}}\right)%
=\sqrt{\frac{(2\ell_{1}+1)(2\ell_{2}+1)(%
2\ell_{3}+1)}{4\pi}}\begin{pmatrix}\ell_{1}&\ell_{2}&\ell_{3}\\
0&0&0\end{pmatrix}\begin{pmatrix}\ell_{1}&\ell_{2}&\ell_{3}\\
m_{1}&m_{2}&m_{3}\end{pmatrix};
\end{equation}
\begin{align}
\label{eq:3j_to_6j}
\begin{pmatrix}j_{1}&j_{2}&j_{3}\\
m_{1}&m_{2}&m_{3}\end{pmatrix}\begin{Bmatrix}j_{1}&j_{2}&j_{3}\\
\ell_{1}&\ell_{2}&\ell_{3}\end{Bmatrix}=\sum_{m^{\prime}_{1}m^{\prime}_{2}m^{\prime}_{3%
}}&(-1)^{\ell_{1}+\ell_{2}+\ell_{3}+m^{\prime}_{1}+m^{\prime}_{2}+m^{\prime}_{3}}\nonumber\\
&\times\begin{pmatrix}j_{1}&\ell_{2}&\ell_{3}\\
m_{1}&m^{\prime}_{2}&-m^{\prime}_{3}\end{pmatrix}\begin{pmatrix}\ell_{1}&j_{2}&\ell_%
{3}\\
-m^{\prime}_{1}&m_{2}&m^{\prime}_{3}\end{pmatrix}\begin{pmatrix}\ell_{1}&\ell_{2}&j_%
{3}\\
m^{\prime}_{1}&-m^{\prime}_{2}&m_{3}\end{pmatrix}.
\end{align}
\end{itemize}

\section{Derivations}
\subsection{Derivation of Eq. (\ref{eqn:legendre_to_spherical})}
\label{app:deri:P_to_Y}
Applying identities (\ref{eq:legen_add}, \ref{eq:sph_sym}, \ref{eq:3j_sym_permu}, \ref{eq:3j_sym_ref}, \ref{eq:Y_to_3j}, \ref{eq:3j_to_6j}), we obtain
\begin{align}
&\mathcal{P}_{\ell}(\hat{\bm{q}}_1\cdot\hat{\bm{q}}_2)\mathcal{P}_{\ell_2}(\hat{\bm{q}}_1\cdot\hat{\bm{k}})\mathcal{P}_{\ell_1}(\hat{\bm{q}}_2\cdot\hat{\bm{k}})\nonumber\\
=&\frac{(4\pi)^3}{(2\ell+1)(2\ell_1+1)(2\ell_2+1)}\sum_{m,m_1,m_2}(-1)^{m+m_1+m_2}\nonumber\\
 &\times Y_{{\ell}m}(\hat{\bm{q}}_1)Y_{{\ell},-m}(\hat{\bm{q}}_2)Y_{\ell_2m_2}(\hat{\bm{q}}_1)Y_{\ell_2,-m_2}(\hat{\bm{k}})Y_{\ell_1m_1}(\hat{\bm{q}}_2)Y_{\ell_1,-m_1}(\hat{\bm{k}})\nonumber\\
=&(4\pi)^{\frac{3}{2}}\sum_{J_1,J_2,J_k}\sqrt{(2J_1+1)(2J_2+1)(2J_k+1)}\left(\begin{array}{ccc}
{\ell}& \ell_2 & J_1\\ 0& 0&0
\end{array}\right)
\left(\begin{array}{ccc}
{\ell}& \ell_1 & J_2\\ 0& 0&0
\end{array}\right)
\left(\begin{array}{ccc}
\ell_2 & \ell_1 & J_k\\ 0& 0&0
\end{array}\right)\nonumber\\
&\times\sum_{M_1,M_2,M_k}(-1)^{M_1+M_2+M_k}Y_{\scriptscriptstyle J_1M_1}(\hat{\bm{q}}_1)Y_{\scriptscriptstyle J_2M_2}(\hat{\bm{q}}_2)Y_{\scriptscriptstyle J_kM_k}(\hat{\bm{k}})\nonumber\\
 &\times \sum_{m,m_1,m_2}(-1)^{m+m_1+m_2}\left(\begin{array}{ccc}
{\ell}& \ell_2 & J_1\\ m& m_2&-M_1
\end{array}\right)
\left(\begin{array}{ccc}
{\ell}& \ell_1 & J_2\\ -m& m_1&-M_2
\end{array}\right)
\left(\begin{array}{ccc}
\ell_2 & \ell_1 & J_k\\ -m_2& -m_1&-M_k
\end{array}\right)\nonumber\\ 
=&(4\pi)^{3/2}(-1)^{\ell_1+\ell_2+{\ell}}\nonumber\\
 &\times\sum_{J_1,J_2,J_k}\sqrt{(2J_1+1)(2J_2+1)(2J_k+1)}\left(\begin{array}{ccc}
J_1 & \ell_2 & \ell\\ 0& 0&0
\end{array}\right)
\left(\begin{array}{ccc}
\ell_1 & J_2 & \ell\\ 0& 0&0
\end{array}\right)
\left(\begin{array}{ccc}
\ell_1 & \ell_2 & J_k\\ 0& 0&0
\end{array}\right)
\left\lbrace\begin{array}{ccc}
J_1 & J_2 & J_k\\ \ell_1 & \ell_2 & \ell
\end{array}\right\rbrace\nonumber\\
 &\times(-1)^{J_1+J_2+J_k}\sum_{M_1,M_2,M_k}Y_{\scriptscriptstyle J_1M_1}(\hat{\bm{q}}_1)Y_{\scriptscriptstyle J_2M_2}(\hat{\bm{q}}_2)Y_{\scriptscriptstyle J_kM_k}(\hat{\bm{k}})\left(\begin{array}{ccc}
J_1 & J_2 & J_k\\ M_1 & M_2 & M_k
\end{array}\right)~,
\end{align}
where we can define a coefficient
\begin{align}
C_{\ell_1\ell_2\ell}^{J_1J_2J_k}\equiv&(4\pi)^{3/2}(-1)^{\ell_1+\ell_2+{\ell}+J_1+J_2+J_k}\nonumber\\
 &\times\sqrt{(2J_1+1)(2J_2+1)(2J_k+1)}\left(\begin{array}{ccc}
J_1 & \ell_2 & \ell\\ 0& 0&0
\end{array}\right)
\left(\begin{array}{ccc}
\ell_1 & J_2 & \ell\\ 0& 0&0
\end{array}\right)
\left(\begin{array}{ccc}
\ell_1 & \ell_2 & J_k\\ 0& 0&0
\end{array}\right)
\left\lbrace\begin{array}{ccc}
J_1 & J_2 & J_k\\ \ell_1 & \ell_2 & \ell
\end{array}\right\rbrace~,
\end{align}
Note that when we combine two spherical harmonics into one, the triangle conditions of the $3j$ symbols imply that
\begin{equation}
M_1= m+m_2~,~~M_2=-m+m_1~,~~M_k=-m_1-m_2~,
\end{equation}
so that $M_1,M_2,M_k$ satisfy
\begin{equation}
M_1+M_2+M_k=0~.
\end{equation}
According to the condition (\ref{eq:even_condition}), we have $J_1+J_2+J_k={\rm even}$, leading to $(-1)^{\scriptscriptstyle J_1+J_2+J_k}=1$.
Hence, Eq. (\ref{eqn:legendre_to_spherical}) is recovered.

\subsection{Derivation of Eq. (\ref{eq:T-r}) and (\ref{eq:a_JJJ})}\label{app:deri:T-r}
Applying Eqs. (\ref{eq:Y_to_3j}) and (\ref{eq:3j_ortho}), we obtain
\begin{align}
&\sum_{M_1M_2}\left(\begin{array}{ccc}
J_1 & J_2 & J_k \\
M_1 & M_2 & M_k
\end{array}\right)Y_{\scriptscriptstyle J_1M_1}(\hat{\bm{r}})Y_{\scriptscriptstyle J_2M_2}(\hat{\bm{r}})\nonumber\\
=&\sum_{M_1M_2}\left(\begin{array}{ccc}
J_1 & J_2 & J_k \\
M_1 & M_2 & M_k
\end{array}\right)\sum_{\ell'm'}\sqrt{\frac{(2J_1+1)(2J_2+1)(2\ell'+1)}{4\pi}}\left(\begin{array}{ccc}
J_1 & J_2 & \ell' \\
M_1 & M_2 & m'
\end{array}\right)Y_{\ell'm'}^*(\hat{\bm{r}})\left(\begin{array}{ccc}
J_1 & J_2 & \ell' \\
0 & 0 & 0
\end{array}\right)\nonumber\\
=&\sum_{\ell'm'}\sqrt{\frac{(2J_1+1)(2J_2+1)}{4\pi(2\ell'+1)}}Y_{\ell'm'}^*(\hat{\bm{r}})\left(\begin{array}{ccc}
J_1 & J_2 & \ell' \\
0 & 0 & 0
\end{array}\right)
\sum_{M_1M_2}(2\ell'+1)\left(\begin{array}{ccc}
J_1 & J_2 & J_k \\
M_1 & M_2 & M_k
\end{array}\right)\left(\begin{array}{ccc}
J_1 & J_2 & \ell' \\
M_1 & M_2 & m'
\end{array}\right)\nonumber\\
=&\sum_{\ell'm'}\sqrt{\frac{(2J_1+1)(2J_2+1)}{4\pi(2\ell'+1)}}Y_{\ell'm'}^*(\hat{\bm{r}})\left(\begin{array}{ccc}
J_1 & J_2 & \ell' \\
0 & 0 & 0
\end{array}\right)\delta_{\scriptscriptstyle \ell'J_k}\delta_{\scriptscriptstyle m'M_k}\nonumber\\
=&\sqrt{\frac{(2J_1+1)(2J_2+1)}{4\pi(2J_k+1)}}Y_{\scriptscriptstyle J_kM_k}^*(\hat{\bm{r}})\left(\begin{array}{ccc}
J_1 & J_2 & J_k\\
0 & 0 & 0
\end{array}\right)~,
\end{align}
where we can define the coefficient as
\begin{equation}
a_{\scriptscriptstyle J_1J_2J_k}\equiv\sqrt{\frac{(2J_1+1)(2J_2+1)}{4\pi(2J_k+1)}}\left(\begin{array}{ccc}
J_1 & J_2 & J_k\\
0 & 0 & 0
\end{array}\right).
\end{equation}
Hence, Eqs. (\ref{eq:T-r}) and (\ref{eq:a_JJJ}) are demonstrated.

\section{Proof of Feasibility of Series Expansion}\label{app:proof}
In this section we will prove the series expansion of $\bar{A}(k,\mu_n)$ and $\bar{B}(k,\mu_n)$ are feasible. Suppose $p_1,p_2$ are non-negative integers, we want to show the following finite series expansion always exists,
\begin{equation}
D(k,\mu_n)\equiv\int\dq{1}(\hat{\bm{q}}_1\cdot\hat{\bm{n}})^{p_1}(\hat{\bm{q}}_2\cdot\hat{\bm{n}})^{p_2} = \sum_{i=0}D_i(k)\mu_n^i~.
\label{eq:mu_expansion}
\end{equation}
In spherical coordinates where the $z-$axis is chosen along $\hat{\bm{k}}$ and $\hat{\bm{n}}$ on the $x-z$ plane, the kernel will be
\begin{align}
F(p_1,p_2)\equiv &(\hat{\bm{q}}_1\cdot\hat{\bm{n}})^{p_1}(\hat{\bm{q}}_2\cdot\hat{\bm{n}})^{p_2}\nonumber\\
=& (\sin\theta_1\sin\theta_n\cos\phi+\cos\theta_1\cos\theta_n)^{p_1}(-\sin\theta_2\sin\theta_n\cos\phi+\cos\theta_2\cos\theta_n)^{p_2}\nonumber\\
=&\sum_{r_1=0}^{p_1}\binom{p_1}{r_1} \sin^{r_1}\theta_1\sin^{r_1}\theta_n\cos^{r_1}\phi\cos^{p_1-r_1}\theta_1\cos^{p_1-r_1}\theta_n\nonumber\\
 &\times\sum_{r_2=0}^{p_2}(-1)^{r_2}\binom{p_2}{r_2}\sin^{r_2}\theta_2\sin^{r_2}\theta_n\cos^{r_2}\phi\cos^{p_2-r_2}\theta_{\theta_2}\cos^{p_2-r_2}\theta_n\nonumber\\
=& \sum_{r_1=0}^{p_1}\sum_{r_2=0}^{p_2}(-1)^{r_2}\binom{p_1}{r_1}\binom{p_2}{r_2}\cos^{r_1+r_2}\phi\sin^{r_1}\theta_1\sin^{r_2}\theta_2\sin^{r_1+r_2}\theta_n\cos^{p_1-r_1}\theta_1\cos^{p_2-r_2}\theta_2\nonumber\\
&~~~~~~~~~~~\times\cos^{p_1+p_2-r_1-r_2}\theta_n~.
\end{align}
Averaging over the azimuthal angle $\phi$, we are only left with terms with $r_1+r_2={\rm even}$, since $\langle\cos^{m}\phi\rangle_{\phi}=0$ for odd integer $m$. The kernel then becomes
\begin{align}
\langle F(p_1,p_2)\rangle=& \sum_{r_1=0}^{p_1}\sum_{r_2=0}^{p_2} (-1)^{r_2}\binom{p_1}{r_1}\binom{p_2}{r_2}\langle\cos^{r_1+r_2}\phi\rangle \sin^{r_1}\theta_1\sin^{r_2}\theta_2\cos^{p_1-r_1}\theta_1\cos^{p_2-r_2}\theta_2\nonumber\\
&~~~~~~~~~~~\times\left(1-\mu_n^2\right)^\frac{r_1+r_2}{2}\mu_n^{p_1+p_2-r_1-r_2}~.
\label{eq:Fp1p2_expansion}
\end{align}
Since $(r_1+r_2)/2$ and $p_1+p_2-r_1-r_2$ are both non-negative integers, we can futher expand it as a polynomial of $\mu_n$. Thus, the expansion (\ref{eq:mu_expansion}) is always feasible.

Furthermore, from Eq. (\ref{eq:Fp1p2_expansion}) we obtain two properties of the expansion:
\begin{enumerate}
\item The power of $\mu_n$ goes up to $p_1+p_2$, so that the series is finite. And it goes as $p_1+p_2-2,p_1+p_2-4,\cdots,$ down to 0 or 1 depending on the parity of $p_1+p_2$.
\item The part with $\cos^{p_1-r_1}\theta_1\cos^{p_2-r_2}\theta_2$ can always be written as products of Legendre polynomials of $\mu_1$ and $\mu_2$. The only apparent problem comes from $\sin\theta_1$ and $\sin\theta_2$. However, since $r_1+r_2$ is even, $r_1-r_2$ must be even as well. Suppose $r_1\geq r_2$, the potentially problematic term becomes:
\begin{eqnarray}
\sin^{r_1}\theta_1\sin^{r_2}\theta_2
&=&
(\sin\theta_1\sin\theta_2)^{r_2}\sin^{r_1-r_2}\theta_1
\nonumber \\
&=&\left(\cos\theta_1\cos\theta_2-\hat{\bm{q}}_1\cdot\hat{\bm{q}}_2\right)^{r_2}\left(1-\cos^2\theta_1\right)^{\frac{r_1-r_2}{2}}~,
\end{eqnarray}
so that each term can be written in terms of the products of $\cos\theta_1,\cos\theta_2$ and $\hat{\bm{q}}_1\cdot\hat{\bm{q}}_2$, which can be further decomposed into Legendre polynomials.
\end{enumerate}







\end{document}